\documentclass[usenatbib]{mn2e}
\usepackage{epsfig,lscape}

\newcommand{\etal}{et~al.}

\newcommand{\kms}{$\mbox{km~s}^{-1}$ }
\newcommand{\kmsns}{$\mbox{km~s}^{-1}$}

\newcommand{\vsfig}[2]           
{
  \begin{center}
    \begin{minipage}[t]{0.05\textwidth}
      {\footnotesize \raisebox{40mm}{(#2)}}
    \end{minipage}
    \begin{minipage}[t]{0.42\textwidth}
      \psfig{file=./#1.ps,height=0.95\textwidth,angle=0}
    \end{minipage}
    \hfill
  \end{center}
}

\newcommand{\specdfig}[2]        
{
   \begin{center}
     \begin{minipage}[t]{0.45\textwidth}
         \psfig{file=eps/#1.eps,width=0.35\textwidth,width=0.8\textwidth,angle=270}
     \end{minipage}
     \hfill
     \begin{minipage}[t]{0.45\textwidth}
         \psfig{file=eps/#2.eps,width=0.35\textwidth,width=0.8\textwidth,angle=270}
     \end{minipage}
   \end{center}
}

\newcommand{\specsfig}[1]        
{
   \begin{center}
     \begin{minipage}[t]{0.45\textwidth}
         \psfig{file=eps/#1.eps,height=0.65\textwidth,width=1\textwidth,angle=0}
     \end{minipage}
   \end{center}
}

\newcommand{\boxfig}[1]        
{
   \begin{center}
     \begin{minipage}[t]{0.46\textwidth}
         \psfig{file=eps/#1.eps,height=0.45\textwidth,angle=0}
     \end{minipage}
   \end{center}
}

\newcommand{\twofig}[2]        
{
   \begin{center}
     \begin{minipage}[t]{0.5\textwidth}
         \psfig{file=eps/#1.eps,height=0.95\textwidth}
     \end{minipage}
     \hfill
     \begin{minipage}[t]{0.5\textwidth}
         \psfig{file=eps/#2.eps,height=0.95\textwidth}
     \end{minipage}
   \end{center}
}

\begin{document}

\title[12.2-GHz catalogue: longitudes 186$^{\circ}$ to 330$^{\circ}$]{12.2-GHz methanol maser MMB follow-up catalogue - II. Longitude range 186$^{\circ}$ to 330$^{\circ}$}
\author[S.\ L.\ Breen \etal]{S.\ L. Breen,$^{1}$\thanks{Email: Shari.Breen@csiro.au} S.\ P. Ellingsen,$^2$ J.\ L. Caswell,$^1$ J.\ A.\ Green,$^1$ M.\ A.\ Voronkov,$^1$ \newauthor G.\ A.\ Fuller,$^3$  L.\ J.\ Quinn,$^3$ A.\ Avison$^3$\\
 \\
  $^1$ CSIRO Astronomy and Space Science, Australia Telescope National Facility, PO Box 76, Epping, NSW 1710, Australia;\\
  $^2$ School of Mathematics and Physics, University of Tasmania, Private Bag 37, Hobart, Tasmania 7001, Australia;\\
  $^3$ Jodrell Bank Centre for Astrophysics, Alan Turing Building, School of Physics and Astronomy, University of Manchester,\\ Manchester M13 9PL, UK}
 
 \maketitle
  
 \begin{abstract}
We present the second portion of a catalogue of 12.2-GHz methanol masers detected towards 6.7-GHz methanol masers observed in the unbiased Methanol Multibeam (MMB) Survey. Using the Parkes radio telescope we have targeted all 207 6.7-GHz methanol masers in the longitude range 186$^{\circ}$ to 330$^{\circ}$ for 12.2-GHz counterparts. We report the detection of 83 12.2-GHz methanol masers, and one additional source which we suspect is thermal emission, equating to a detection rate of 40 per cent. Of the 83 maser detections, 39 are reported here for the first time. We discuss source properties, including variability and highlight a number of unusual sources. We present a list of 45 candidates that are likely to harbor methanol masers in the 107.0-GHz transition. 
\end{abstract}


\section{Introduction}

Masers are key tools for investigating the formation of high-mass stars. They are not only sensitive probes for the discovery of very young high-mass star formation regions, but also for the kinematics of the regions that they are associated with. Furthermore, they are relatively common, strong and arise from many chemical species and transitions. Some of these transitions are very common, like the 22-GHz transition of water, the 6.7- and 12.2-GHz transitions of methanol and the 1.6-GHz hydroxyl transitions, while others are much rarer, such as the 37.7-GHz and 107.0-GHz methanol masers \citep[e.g.][]{Ellingsen11}. Since the different masers are created under very specific sets of physical conditions \citep[e.g.][]{Cragg05}, the detection of the different combinations of masers can reveal details of the physical conditions in these regions, and potentially the evolutionary stage of the associated star formation region \citep[e.g.][]{Ellingsen07}.

The Methanol Multibeam (MMB) survey is a project that aims to search the entire Galactic plane (within latitudes of $\pm$2$^{\circ}$) for 6.7-GHz methanol masers \citep{Green09}. Methanol masers at 6.7-GHz are especially useful as they exclusively trace sites of high-mass star formation \citep[e.g.][]{Minier03,Xu08} and trace the systemic velocities \citep[e.g.][]{Szy07,Pandian09,GM11} of the regions that they are associated with, making them excellent tools for investigating not only the kinematics and the physical conditions of these regions, but also aspects like Galactic structure \citep[e.g.][]{Green11}. The southern hemisphere component of the survey has been completed using the Parkes radio telescope and the survey results have now been published in the longitude range 186$^{\circ}$ (through the Galactic centre) to 20$^{\circ}$ \citep{CasMMB10,GreenMMB10,CasMMB102,Green12}. In this longitude range 707 6.7-GHz methanol masers have been detected, about 40 per cent of which were new to the survey.

We have conducted 12.2-GHz observations towards all 6.7-GHz methanol masers detected in the southern portion of the MMB survey, constituting the largest, statistically robust sample of 12.2-GHz masers. This large, sensitive search has allowed us to overcome the biases present in previous studies of these masers. The first portion of the 12.2-GHz MMB methanol maser follow-up catalogue covering the longitude range 330$^{\circ}$ (through 360$^{\circ}$) to 10$^{\circ}$ \citep{BreenMMB12} presented 184 12.2-GHz methanol masers, 117 of which were reported for the first time. The global properties of all 12.2-GHz methanol masers associated with MMB sources that lie south of declination --20$^{\circ}$ (Galactic longitudes $\sim$254$^{\circ}$ (through 360$^{\circ}$) to $\sim$10$^{\circ}$) have also been presented \citep{Breen12stats} and used two of the three epochs of data presented here (2008 June and December). 

Give that there have been no serendipitous discoveries of 12.2-GHz methanol masers without accompanying 6.7-GHz emission, combined with the rarity of 12.2-GHz emission surpassing the flux density of associated 6.7-GHz emission \citep[e.g.][]{Caswell95b,Blas04,Gay,BreenMMB12}, it is likely that the 12.2-GHz masers detected in our search represents a complete sample of these masers. Since the physical conditions required to produce these two methanol maser transitions are similar \citep[e.g][]{Cragg05}, observations of a large number of 6.7-GHz methanol masers with and without associated 12.2-GHz are especially useful probes of the physical conditions in their immediate vicinities.


Here we present the second portion of the 12.2-GHz catalogue, covering a longitude range of 186$^{\circ}$ to 330$^{\circ}$, wholly matching the longitude range of the MMB catalogue presented in \citet{Green12}, and including 207 6.7-GHz methanol maser targets. In addition to the usual $\pm$2$^{\circ}$ latitude coverage of the MMB survey, this longitude range also includes observations of the Orion-Monoceros region. The remaining 12.2-GHz sources that we have observed (longitudes $>$10$^{\circ}$) are to be published following the publication of the MMB targets.

\section{Observations and data reduction} 

The Parkes 64-m radio telescope was used to conduct follow-up observations at 12.2-GHz towards MMB sources in the longitude range 186$^{\circ}$ to 330$^{\circ}$. The  12.2-GHz observing sessions; 2008 June 20-25, 2008 December 5-10 and 2010 March 19-23, were carried out as close in time as could be achieved to the 6.7-GHz MMB methanol maser spectra (which were mainly taken during 2008 March and August and 2009 March). This minimises the effects of source variability, thereby allowing us to meaningfully compare source properties. The 6.7-GHz targets have been positioned with subarcsec uncertainties, and we expect the positions of the two transitions to be in good agreement in accordance with results from a small sample for which precise comparisons have been made \citep[e.g.][]{Norris93,Mos02,Goed05a}.

 \citet{Breen12stats} detailed the observing strategy and specifications used for the observations, as briefly summarised here. The targeted 12.2-GHz observations were made using the {\em Ku}-band receiver on the Parkes radio telescope, which detected two orthogonal linear polarisations. The system equivalent flux densities were 205 and 225~Jy for the respective polarizations during observations in 2008 June, and slightly higher at 220 and 240~Jy during 2008 December and 2010 March. The Parkes multibeam correlator was configured to record 8192 channels over 16-MHz for each polarisations, resulting in a usable velocity coverage of $\sim$290 \kms and a spectral resolution of 0.08 \kmsns, after Hanning smoothing which was applied during the data processing. The Parkes radio telescope has rms pointing errors of $\sim$10 arcsec and at 12.2 GHz the telescope has a half power beam width of 1.9 arcmin. Flux density calibration was with respect to PKS B1934--638, which has an assumed flux density of 1.825~Jy at 12178-MHz \citep{Sault03}, and was observed daily. The system was stable over the several days of each observing period and we expect our absolute flux density calibration to be accurate to $\sim$10 per cent. 

All sources were observed at a single frequency of 12178~MHz (i.e. with no Doppler tracking), allowing us to forego the traditional observing mode which requires unique reference spectra to be obtained for each on-source observation. We created a sensitive reference bandpass made up of the median value of each spectral channel from all of the spectra collected during each of the observing sessions. This method effectively reduces the required telescope time by half, and also results in lower noise in the quotient spectrum since the reference bandpass has a long effective integration time and thus adds little noise. This strategy introduced a baseline ripple that we have removed by subtracting a running median over 100 channels from the final quotient spectrum. This method is best suited to the narrow maser lines that we detect, but makes it difficult to reliably recover detected absorption features since they can be comparable to the ripple in both channel width and amplitude. A further complication can arise when very strong maser emission is present over a wide velocity range, causing small negative dips to flank the emission. This artefact was investigated in the first portion of the catalogue \citep{BreenMMB12}, and appeared to only affect very strong sources such as G\,9.621+0.196 and we do not believe that any of the spectra presented in this longitude range have been distorted in this manner. 

Integration times for each of the maser targets vary from a total of 5 to $\sim$20 minutes over the three observation epochs. 12.2-GHz non-detections were routinely observed during at least two of the three Parkes sessions, with a minimum total of 10 minutes on source. Data were reduced using the ATNF (Australia Telescope National Facility) Spectral Analysis Package (ASAP) and included alignment of velocity channels which are given with respect to LSR. The adopted rest frequency was 12.178597 GHz \citep{Muller04}. The resultant 5-$\sigma$ detection limits for each source in a single observing session, after averaging the two polarisations, chiefly lie in the range 0.55 to 0.80~Jy, which is comparable to the 5-$\sigma$ sensitivity of the MMB survey \citep[0.85~Jy;][]{Green09}. After inspecting each epoch of data separately for detected maser emission, multiple epochs were averaged and further inspected for maser emission, allowing us to identify several additional weak sources. 

No observation of G\,212.06--0.74 was made during any of the Parkes observing epochs. For this source we present data taken with the University of Tasmania Hobart 26-m radio telescope (during 2012 April), made using a cryogenically cooled receiver that, at the time of the observation, detected a single circularly polarised signal. The system equivalent flux density during the observations was $\sim$900~Jy. The data were recorded with 4096 channels over a bandwidth of 8-MHz, yielding the same channel spacing and therefore spectral resolution as achieved in the Parkes observations. An integration time of two hours was used and a unique reference observation was made, resulting in an rms noise level of 0.34~Jy. Absolute flux density calibration was with respect to Virgo A, which we assumed had a flux density of 35~Jy \citep{Ott94}, and is expected to be accurate to $\sim$30 per cent.

\section{Results}\label{sect:results}

Our 12.2-GHz observations targeting the 207 6.7-GHz methanol masers (listed in Table~\ref{tab:6MMB}) presented by \citet{Green12} in the longitude range 186$^{\circ}$ to 330$^{\circ}$ have resulted in the detection of 83 12.2-GHz methanol masers. We interpret the emission detected towards G\,208.996--19.386 as thermal. The detection rate over this longitude range is slightly lower than found in the 330$^{\circ}$ (through 360$^{\circ}$) to 10$^{\circ}$ longitude region \citep{BreenMMB12}, at 40 per cent (compared to 46 per cent). About half the detected sources (39 of 83) are new discoveries. References to the previously detected 12.2-GHz methanol masers are given in Table~\ref{tab:12MMB}.

The characteristics of the target 6.7-GHz methanol masers are listed in Table~\ref{tab:6MMB}, together with the results of our 12.2-GHz methanol maser observations. For some of the analysis undertaken (especially in \citet{Breen12stats}) the integrated flux density data for both the 6.7 and 12.2-GHz transitions were required. The former was not presented in the MMB survey papers \citep{CasMMB10,GreenMMB10, CasMMB102,Green12} and we have extracted this information from the MMB `MX' spectra. `MX' refers to an observational mode used in MMB follow-up observations whereby a source was tracked, and the pointing centre cycled through each of the receiver's seven beams. For internal consistency we also independently determined the peak flux density and velocity range of the 6.7-GHz emission. There are only minor discrepancies between our values and those reported in the MMB catalogues and these have most likely arisen from using data from slightly different epochs. The outcomes of the 12.2-GHz observations are also listed, noting either `detection' or the 5-$\sigma$ detection limits for non-detections. Where 12.2-GHz observations were completed during two or more epochs, the data-averaged rms noise is usually a factor of $\sqrt{2}$ or more better than those which are listed. Sources requiring further comments to understand their complicated nature or that are of special interest are marked with an `*' and have further details presented in Section~\ref{sect:ind}.

Table~\ref{tab:12MMB} presents the characteristics of the 12.2-GHz sources that we detect. 12.2-GHz masers have been reliably identified at flux densities lower (3- to 4-$\sigma$) than the traditional 5-$\sigma$ detection limits when the emission perfectly corresponds in velocity with the target 6.7-GHz sources. The details presented in Table~\ref{tab:12MMB} follow the usual practice in column 1 where the Galactic longitude and latitude is used as a source name for each source (references to previous 12.2-GHz observations are noted as superscripts after source names);  columns 2 and 3 give the source right ascension and declination of the 6.7-GHz measured position; column 4 gives the epoch of the 12.2-GHz observations that are referred to in columns 5 to 8 which give the peak flux density (Jy), velocity of the 12.2-GHz peak (\kmsns), velocity range (\kmsns) and integrated flux density (Jy \kmsns), respectively.

Fig.~\ref{fig:12MMB} presents spectra of each of the detected 12.2-GHz methanol masers. Sources are shown in order of increasing Galactic longitude except where vertical alignment of spectra corresponding to nearby sources was necessary to highlight where features of a source were also detected at nearby positions. The epoch of the observation, either 2008 June, 2008 December or 2010 March, is annotated in the top left hand corner of each of the spectra. For some weak sources the epoch-averaged spectrum is presented and these are annotated with the two or three observing dates. The velocity range of the majority of spectra is 30 \kmsns, usually centred on the velocity of the 6.7-GHz methanol maser peak emission, giving an immediate feeling for the basic structure of the two methanol maser transitions; and, in a number of cases highlights that lone features detected at 12.2-GHz are associated with the 6.7-GHz peak emission. For some sources, the centre velocity of the spectra had to be changed from the 6.7-GHz peak velocity in order to contain the full extent of the 6.7-GHz emission, this was especially common for sources that were complicated by emission from nearby sources and therefore had to be aligned. Sources not centred on the velocity of the 6.7-GHz peak are;  G\,305.199+0.005, G\,305.200+0.019, G\,305.202+0.208, G\,305.208+0.206, G\,316.381--0.379, G\,328.237--0.547, G\,328.254--0.532 and G\,329.407--0.459.

\begin{table*}\footnotesize
 \caption{Characteristics of the 6.7-GHz methanol maser targets as well as a brief description of the 12.2-GHz results (either detection or 5-$\sigma$ detection limits). The full complement of 12.2-GHz source properties are listed in Table~\ref{tab:12MMB}. Column 1 gives the Galactic longitude and latitude of each source and is used as an identifier (a `*' indicates sources with notes in Section~\ref{sect:ind}; and a `$^{sc}$' following the names G\,294.997--1.734 and G\,321.704+1.168 indicates that its 6.7-GHz properties have been extracted from the survey cube rather than follow-up `MX' observations); column 2 gives the peak flux density (Jy) of the 6.7-GHz sources, derived from follow-up MX observations at the accurate 6.7-GHz position unless otherwise noted; columns 3 and 4 give the peak velocity and the minimum and maximum velocity (\kmsns) of the 6.7-GHz emission respectively (also derived from Parkes MX observations); column 5 gives the integrated flux density of the 6.7-GHz emission (Jy \kmsns). A `--' in either of the detection limit columns (i.e. columns 6, 8 or 10) indicates that no observations were made on the given epoch. The values listed in columns 6-11 are replaced with the word `detection' where 12.2-GHz emission is observed (a $^{\gamma}$ following `detection' for G\,208.996--19.386 indicates that we interpret the emission as being thermal). The presence of `conf' in the columns showing the integrated flux density indicates that a value for this characteristic could not be extracted do to confusion from nearby sources. Columns 6-11 give the 5-$\sigma$ 12.2-GHz methanol maser detection limits and observed velocity ranges for the 2008 June, 2008 December and 2010 March epochs respectively.}
  \begin{tabular}{lllllclllclllcllllll} \hline
 \multicolumn{1}{c}{\bf Methanol maser} &\multicolumn{4}{c}{\bf 6.7-GHz properties} & \multicolumn{6}{c}{\bf 12.2-GHz observations}\\
    \multicolumn{1}{c}{\bf ($l,b$)}& {\bf S$_{6.7}$} & {\bf Vp$_{6.7}$} & {\bf Vr$_{6.7}$ } & {\bf  I$_{6.7}$} & \multicolumn{2}{c}{\bf 2008 June} & \multicolumn{2}{c}{\bf 2008 December} & \multicolumn{2}{c}{\bf 2010 March}\\	
      \multicolumn{1}{c}{\bf (degrees)}  &{\bf (Jy)} &&{\bf (\kmsns)}& & \multicolumn{1}{c}{\bf 5-$\sigma$} & \multicolumn{1}{c}{\bf Vr} & \multicolumn{1}{c}{\bf 5-$\sigma$} & \multicolumn{1}{c}{\bf Vr} & \multicolumn{1}{c}{\bf 5-$\sigma$} & \multicolumn{1}{c}{\bf Vr}\\  \hline \hline
G\,188.794+1.031	&	9.6	&	--5.4	&	--5.7,--4.5	& 4.3		&$<$0.75	& --163,125		& --&  & --	\\ 
G\,188.946+0.886*	&	607	&	10.8	&	7.6,12.1	&444	& \multicolumn{6}{c}{detection}\\ 
G\,189.030+0.783	&	17	&	8.9	&	8.6,10.9		&	8.2	&	$<$0.77 & --210,130 &  $<$0.53  & --190,150 & -- \\ 
G\,189.471--1.216	&	1.9	&	18.8	&	18.6,18.8	&	0.4	&-- & &-- &  &  $<$0.79 & --220,120 	\\
G\,189.778+0.345	&	4.6	&	5.5	&	3.6,5.9	&	2.5	&	$<$0.75	&	--200,140 & $<$0.53	& --190,150 & --\\
G\,192.600--0.048	& 	83	&	4.6	&	1.6,6.0	&	64 &    \multicolumn{6}{c}{detection}\\ 
G\,196.454--1.677*	&	22	&	15.2	&	14.5,16.0	&	16	&	$<$0.73	& 	--200,140	& $<$0.53	& --190,150 & --\\
G\,206.542--16.355	&	1.0	&	12.3	&	12.0,12.4	& 0.4 	&--		&	& $<$0.83		&  --225,115 &   $<$0.58	& --215,125& \\ 
G\,208.996--19.386*	&$<$0.2	&	7.3	&	5.0,10.0	&		&	\multicolumn{6}{c}{detection$^{\gamma}$}\\  
G\,209.016--19.398	&	1.4	&	--1.5	&	--1.9,--1.3 & 0.6	&-- & & 	$<$0.99	&	--220,120	&  $<$0.69 &	--220,120	&	\\
G\,212.06--0.74*	&	0.3	&	44.3	&	42.8,49.8	&	0.6	&	$<$1.7 &   --60,100 & --&  &-- &   &\\
G\,213.705--12.597	&	292	&	10.8	&	9.9,13.8	&	171	&	\multicolumn{6}{c}{detection}\\
G\,232.620+0.996	&	165	&	22.9	&	21.4,23.8	&	87	&	\multicolumn{6}{c}{detection}\\
G\,254.880+0.451	&	1.6	&	30.2	&	30.0,30.5	&	0.5	&	$<$0.75		&	--210,130	&		$<$1.35	&	--205,135	& --	\\
G\,259.939--0.041	&	1.4	&	--1.0	&	--1.2,--0.9	&	0.3	&	$<$0.50	&	--210,130	&		--	&			& $<$0.77		&	--200,140&\\
G\,263.250+0.514	&	67	&	12.3	&	11.7,16.9	&	60	&	\multicolumn{6}{c}{detection}\\
G\,264.140+2.018	&	7.6	&	8.3	&	7.9,8.7	&	2.2	&	--	&		&	$<$0.60	&			--205,135	&			$<$0.57	&	--200,140\\ 
G\,264.289+1.469	&	0.5	&	8.7		&	8.5,8.8	&	0.2	&	\multicolumn{6}{c}{detection}\\
G\,269.153--1.128	&	4.1	&	16.0		&	8.8,16.3	&	1.9	&	$<$0.75	&	--210,130	&		--	&			& 	$<$0.81	&	--200,140\\ 
G\,269.456--1.467	&	5.2	&	56.1		&	53.7,56.5	&	2.2	&	\multicolumn{6}{c}{detection}\\
G\,269.658--1.270	&	5.4	&	16.3		&	14.2,16.8	&	3.1	&	$<$0.50	&	--210,130	&	$<$0.75	&	--205,135	&	 --\\
G\,270.255+0.835	&	0.4	&	4.0		&	3.0,5.2		&	0.3	&	$<$0.75	&	--210,130	&		--	&			&	 $<$0.79		&	--200,140\\
G\,281.710--1.104	&	0.3	&	1.2		&	1.0,1.2		&	0.1	&	$<$0.55	&	--210,130	&	$<$0.55	&	--205,135	&		--\\ 
G\,284.352--0.419*	&	2.5	&	4.0		&	2.7,4.5	&	1.1	&	$<$0.75	&	--210,130	&	$<$0.55	&	--205,135	&	--	\\
G\,284.694--0.361	&	3.3	&	13.4		&	12.1,13.7	&	1.8	&	$<$0.55	&	--210,130	&	$<$0.75	&	--205,135	&	 --	\\
G\,285.337--0.002	&	12	&	0.8		&	--8.2,2.5	&	11	&	\multicolumn{6}{c}{detection}\\
G\,286.383--1.834	&	16	&	9.6		&	8.3,10.1	&	7.7	&	\multicolumn{6}{c}{detection}\\
G\,287.371+0.644	&	104	&	--1.8		&	--2.7,--0.3	&	69	&	\multicolumn{6}{c}{detection}	\\
G\,290.374+1.661			&	2.1	&	--24.1	&--26.4,--23.0	&	1.1	&	$<$0.80	&	--205,135	&	$<$0.55	&	--200,140	& --	\\
G\,290.411--2.915			&	4.6	&	--16.0		&--16.9,--14.6	&	2.2	&	\multicolumn{6}{c}{detection}\\
G\,291.270--0.719			&	7.4	&	--31.1	&--34.4,--25.8	&	2.5	&	$<$0.75	&	--205,135	&	$<$0.90	&	--200,140	&	--\\
G\,291.274--0.709			&	65	&	--30.5	&--31.0,--28.0	&	89	&	\multicolumn{6}{c}{detection}\\
G\,291.579--0.431			&	1.0	&	15.3		&	11.7,19.3	&	1.2	&	\multicolumn{6}{c}{detection}	\\ 
G\,291.582--0.435			&	2.9	&	10.4	&	9.3,10.7	&	1.7	&	$<$0.55		&	--200,140	&	--		&				&$<$0.78		& --180,160	\\
G\,291.642--0.546			&	0.2	&	12.1	&	12.1,12.2	&	0.1	&	$<$0.60		&	--180,160	&	$<$0.85	&	--200,140		& --	\\
G\,291.879--0.810			&	1.2	&	33.6	&	30.5,34	&	0.7	&	$<$0.55		&	--180,160	&	$<$0.75	&	--200,140		& --	\\
G\,292.074--1.131*			&	0.8	&	--19.0	&--19.1,--18.9	&	0.2	&	$<$0.55		&	--180,160	&	$<$0.75	&	--200,140		& --	\\ 
G\,292.468+0.168			&	4.3	&	10.4	&	8.3,11.5	&	4.5	&	$<$0.50		&	--180,160	&	$<$0.75	&	--200,140		& --\\
G\,293.723--1.742			&	0.6	&	24.4	&	23.2,25.5	&	0.2	&	\multicolumn{6}{c}{detection}\\ 
G\,293.827--0.746			&	2.5	&	36.9	&	35.0,39.2	&	2.9	&	$<$0.55		&	--200,140	&		--	&				&	$<$0.78	&	--180,160\\
G\,293.942--0.874			&	4.3	&	41.1	&	37.4,41.3	&	1.3	&	\multicolumn{6}{c}{detection}\\
G\,294.337--1.706			&	0.1	&--10.6	&--11.9,--9.1	&	0.1	&	$<$0.50		&	--180,160	&	$<$0.75	&	--200,140		&	 --\\
G\,294.511--1.621			&	7.1	&--11.9	&--13.3,--5.4	&	15	&	$<$0.55		&	--200,140	&	--		&				& $<$0.78	& --180,160\\
G\,294.977--1.734$^{sc}$		&	1.8	&--5.3	&--5.9,--5.0	&	1.1	&	$<$0.50		&	--180,160	&	$<$0.75	&	--200,140		& --\\
G\,294.990--1.719			&	11	&--12.2	&--12.6,--11.8	&	4.3	&	$<$0.55		&	--200,140	&	--		&				&$<$0.77	&	 --180,160 \\
G\,296.893--1.305			&	1.2	&	22.2	&	21.0,22.9	&	1.4	&	$<$0.55		&	--180,160	&	--		&				& $<$0.76		& --180,160&\\
G\,297.406--0.622			&	1.4	&	27.8	&	27.6,28.1	&	0.5	&	$<$0.55		&	--180,160	&	$<$0.75	&	--200,140		& --\\
G\,298.177--0.795			&	2.6	&	23.5	&	22.9,27.8	&	3.3	&	$<$0.55		&	--175,165	&	$<$0.75	&	--200,140		& --	\\
G\,298.213--0.343			&	2.4	&	33.4	&	32.7,38.0	&	2.3	&	$<$0.55		&	--175,165	&	$<$0.80	&	--200,140		& --	\\
G\,298.262+0.739			&	15	&--30.1	&--30.9,--29.1	&	9.3	&	$<$0.55		&	--200,140	&	--		&				& $<$0.79		& --180,160	\\
\end{tabular}\label{tab:6MMB}
\end{table*}

\begin{table*}\addtocounter{table}{-1}
  \caption{-- {\emph {continued}}}
  \begin{tabular}{lllllclllclllcll} \hline
 \multicolumn{1}{c}{\bf Methanol maser} &\multicolumn{4}{c}{\bf 6.7-GHz properties} & \multicolumn{6}{c}{\bf 12.2-GHz observations}\\
    \multicolumn{1}{c}{\bf ($l,b$)}& {\bf S$_{6.7}$} & {\bf Vp$_{6.7}$} & {\bf Vr$_{6.7}$ } & {\bf  I$_{6.7}$} & \multicolumn{2}{c}{\bf 2008 June} & \multicolumn{2}{c}{\bf 2008 December} & \multicolumn{2}{c}{\bf 2010 March}\\	
      \multicolumn{1}{c}{\bf (degrees)}  &{\bf (Jy)} && {\bf (\kmsns)}& & \multicolumn{1}{c}{\bf 5-$\sigma$} & \multicolumn{1}{c}{\bf Vr} & \multicolumn{1}{c}{\bf 5-$\sigma$} & \multicolumn{1}{c}{\bf Vr} & \multicolumn{1}{c}{\bf 5-$\sigma$} & \multicolumn{1}{c}{\bf Vr}\\  \hline \hline
G\,298.632--0.362			&	1.3	&	37.3		&	37.0,45.2		&	1.0	&	\multicolumn{6}{c}{detection}\\ 
G\,298.723--0.086			&	1.1	&	23.5		&	13.8,24.6		&	0.5	&	$<$0.50		&	--175,165	&	$<$0.75	&	--200,140		& --	\\
G\,299.013+0.128			&	8.2	&	18.4		&	18.1,19.1		&	3.3	&	$<$0.55		&	--200,140	&	$<$0.75	&	--200,140		& --	\\
G\,299.772--0.005*			&	16	&	--6.8		&	--9.6,--0.2		&	13	&\multicolumn{6}{c}{detection}	\\ 
G\,300.504--0.176			&	4.1	&	7.4		&	2.9,10.1		&	4.4	&	$<$0.55		&	--200,140	&		--	&				& $<$0.78		& --180,160\\
G\,300.969+1.148			&	4.7	&--37.1		&--39.4,--35.4		&	5.5	&\multicolumn{6}{c}{detection}\\
G\,301.136--0.226			&	1.7	&--39.7		&--40.2,--37.8		&	1.5	&	$<$0.55		&	--175,165	&		--	&				& $<$0.77		&	--180,160	\\ 
G\,302.032--0.061			&	11	&--35.5		&--42.6,--33.8		&	14	&	$<$0.55		&	--200,140 &	--		&				& $<$0.8	 	& --170,170	\\ 
G\,302.034+0.625			&	1.3	&--39.1		&--46.9,--38.9		&	1.3	&	--			&			&	$<$0.55	&	--200,140		& $<$0.8		& --170,170	\\	
G\,302.455--0.741			&	1.2	&	32.6		&	32.4,38.0		&	1.1	&	--			&			&	$<$0.55	&	--200,140		& $<$0.75		&  --170,170\\
G\,303.507--0.721			&	2.1	&	14.2		&	10.7,17.7		&	1.2	&	--			&			&	$<$0.55	&	--200,140		& $<$0.8		&	--170,170\\
G\,303.846--0.363			&	7.4	&	24.8		&	22.5,32.9		&	8.6	&	\multicolumn{6}{c}{detection}\\ 
G\,303.869+0.194			&	0.9	&--36.9		&--37.1,--36.2		&	0.6	&	$<$0.50		&	--175,165	&	$<$0.75	&	--200,140		&	 --\\
G\,304.367--0.336			&	0.9	&	32.7		&	31.9,40.5		&	0.5	&	$<$0.75		&	--175,165	&	$<$0.75	&	--200,140		& --\\
G\,304.887+0.635			&	1.0	&--35.1		&--35.7,--34.9		&	0.3	&	$<$0.80		&	--175,165	&	$<$0.75	&	--200,140	&	& --	\\
G\,305.199+0.005			&	6.0	&--42.7		&--44.3,--38.0		&	7.8	&	\multicolumn{6}{c}{detection}\\ 
G\,305.200+0.019			&	45	&--33.1		&--37.6,--30.3		&	45	&	\multicolumn{6}{c}{detection}\\
G\,305.202+0.208*			&	91	&--44.0		&--46.1,--43.0		&	41	&	\multicolumn{6}{c}{detection}	\\
G\,305.208+0.206*			&	457	&--38.3		&--43.0,--33.4		&	375	&	\multicolumn{6}{c}{detection}	\\
G\,305.248+0.245			&	7.1	&--32.2		&--35.2,--28.2		&	7.3	&	$<$0.55	&	--200,140	&	--	&	& $<$0.80		&	--170,170\\
G\,305.362+0.150			&	5.0	&--36.4		&--37.2,--34.8		&	3.5	&	$<$0.55	&	--200,140	&	--	&	& $<$0.80		&	--170,170	\\ 
G\,305.366+0.184*			&	3.2	&--33.8		&--47.2,--30.0		&	3.0	&	\multicolumn{6}{c}{detection}	\\
G\,305.475--0.096			&	2.9	&--35.4		&--53.9,--27.0		&	5.6	&	$<$0.80	&	--175,165	&	$<$0.80		&	--200,140	& --	\\
G\,305.563+0.013			&	4.9	&	--37.3	&	--41.4,--32.3	&	6.7	&	\multicolumn{6}{c}{detection}	\\
G\,305.573--0.342			&	1.0	&	--51.0	&	--54.6,--50.5	&	0.7	&	--	&			&	$<$0.55	&		--200,140	&		$<$0.80	& 	 	--170,170\\
G\,305.615--0.344			&	3.6	&	--34.9	&	--35.5,--26.6	&	7.2	&	$<$0.75		&	--175,165	&	$<$0.55		&	--200,140	&	--\\
G\,305.634+1.645			&	7.2	&	--54.8	&	--57.5,--54.4	&	4.8	&	$<$0.75	&		--175,165	&			$<$0.75	&			--200,140	&			--\\
G\,305.646+1.589			&	3.9	&	--58.1	&	--59.6,--54.7	&	4.0	&	$<$0.75		&	--175,165	&		$<$0.75		&	--200,140			&	--\\
G\,305.799--0.245			&	0.5	&	--36.4	&	--39.8,--27.1	&	0.5	&	$<$0.55	&	--200,140	&		--	&		&	$<$0.75	&	--170,170\\
G\,305.822--0.115			&	2.7	&	--42.1	&	--46.5,--38.6	&	1.9	&	$<$0.75	&			--175,165			&	$<$0.75			&	--200,140		&	--\\
G\,305.887+0.017			&	9.5	&	--34.1	&	--34.7,--29.5	&	4.5	&	$<$0.80	&			--175,165	&			$<$0.75	&			--200,140		&	--\\
G\,305.940--0.164			&	0.8	&	--51.0	&	--51.1,--50.8	&	0.3	&	\multicolumn{6}{c}{detection}	\\
G\,306.322--0.334			&	0.6	&	--24.6	&	--24.6,--22.4	&	0.4	&	$<$0.50			&	--200,140			&	--	&		&	$<$0.80	&	--170,170\\
G\,307.132--0.476			&  	1.2	&	--33.9	&	--34.9,--33.7	&	0.6	&	$<$0.75	&			--175,165			&	$<$0.75			&	--200,140		&	--	\\
G\,307.133--0.477			&	2.4	&	--38.7	&	--39.7,--36.6	&	2.8	&	$<$0.75	&			--175,165			&	$<$0.75			&	--200,140		&	--	\\
G\,308.056--0.396			&	1.5	&	--11.8	&	--12.0,--10.3	&	0.5	&	$<$0.75	&			--175,165			&	$<$0.75			&	--200,140			&	--\\
G\,308.075--0.411			&	0.9	&	--7.4		&	--7.5,--7.0		&	0.3	&	$<$1.0	&			--175,165			&	$<$1.0			&	--200,140			&	--\\ 
G\,308.651--0.507			&	5.7	&	3.1		&	1.6,3.4		&	4.4	&	\multicolumn{6}{c}{detection}	\\
G\,308.686+0.530			&	1.8	&	--52.9	&	--57.7,--44.1	&	1.4	&	$<$0.75	&		--175,165	&		$<$0.75		&	--200,140	&		--\\
G\,308.715--0.216			&	1.0	&	--12.6	&	--16.6,--9.7	&	1.5	&	$<$0.55	&		--175,165	&		$<$0.75	&		--200,140		&	--\\
G\,308.754+0.549			&	12	&	--45.3	&	--51.4,--38.5	&	19	&	\multicolumn{6}{c}{detection}	\\
G\,308.918+0.123*			&	42	&	--54.7	&	--55.7,--52.4	&	44	&	\multicolumn{6}{c}{detection}	\\
G\,309.384--0.135			&	1.4	&	--49.7	&	--50.3,--49.5	&	0.6	&	$<$0.50		&	--175,165	&	$<$0.75		&	--200,140	&			--\\
G\,309.901+0.231			&	21	&	--54.6	&	--56.6,--52.0	&	17	&	\multicolumn{6}{c}{detection}	\\
G\,309.921+0.479			&	885	&	--59.7	&	--64.1,--53.8	&	1038	&	\multicolumn{6}{c}{detection}	\\
G\,310.144+0.760			&	78	&	--55.8	&	--58.8,--53.9	&	99	&	\multicolumn{6}{c}{detection}	\\
G\,310.180--0.122			&	1.1	&	3.2		&	3.0,3.9		&	0.7	&	$<$0.50	&			--170,170	&			$<$0.75	&			--200,140	&		 --\\
G\,311.230--0.032			&	2.3	&	24.8		&	22.8,26.6		&	1.9	&	$<$0.75	&			--170,170	&			$<$0.75	&			--200,140	&		 --\\
G\,311.551--0.055			&	1.0	&	--56.2	&	--57.8,--56.1	&	0.4	&	\multicolumn{6}{c}{detection}	\\
G\,311.628+0.266			&	4.3	&	--57.6	&	--61.3,--55.2	&	5.0	&	\multicolumn{6}{c}{detection}	\\
G\,311.643--0.380			&	10	&	32.6		&	31.5,36.1		&	8.8	&	$<$0.80	&		--170,170	&			$<$0.75	&			--200,140	&			--\\
G\,311.729--0.735			&	0.5	&	31.1		&	25.7,31.2		&	0.3	&	$<$0.75			&	--170,170			&	$<$0.75			&	--200,140			&	--\\
G\,311.947+0.142			&	0.3	&--43.9		&	--47.2,--43.7	&	0.1	&	$<$0.75			&	--170,170			&	$<$0.75			&	--200,140		&	--	\\
G\,312.071+0.082			&	68	&--34.8		&	--35.3,--29.9	&	76	&	\multicolumn{6}{c}{detection}	\\
G\,312.108+0.262			&	20	&--49.9		&	--54.1,--48.4	&	16	&	$<$0.75			&	--170,170			&	$<$0.75			&	--200,140		& --	\\
G\,312.307+0.661			&	4.1	&--12.3		&	--12.7,--11.9	&	1.5	&	$<$0.75			&	--170,170			&	$<$0.75			&	--200,140		& --	\\
G\,312.501--0.084			&	1.2	&	21.8		&	21.5,24.7		&	0.5	&	$<$0.75			&	--170,170			&	$<$0.75			&	--200,140			& --\\
G\,312.597+0.045			&	0.9	&--59.6		&	--59.9,--59.3	&	0.4	&	$<$0.75			&	--170,170			&	$<$0.75			&	--200,140		& --	\\
G\,312.598+0.045			&	23	&--67.8		&	--68.2,--64.1	&	8.3	&	$<$0.75			&	--170,170			&	$<$0.75			&	--200,140		& --	\\
G\,312.698+0.126			&	1.6	&29.5		&	28.1,35.4		&	1.1	&	$<$0.75			&	--170,170			&	$<$0.75			&	--200,140		& --	\\
G\,312.702--0.087			&	0.8	&--59.3		&	--59.3,--55.5	&	0.9	&	$<$0.80			&	--170,170			&	$<$0.75			&	--200,140			&--\\
G\,313.469+0.190			&	30	&--9.3		&	--15.5,--5.1	&	36	&	\multicolumn{6}{c}{detection}	\\
G\,313.577+0.325			&	101	&--47.8		&	--53.3,--46.0	&	56	&	\multicolumn{6}{c}{detection}	\\ 
\end{tabular}\label{tab:6MMB}
\end{table*}

\begin{table*}\addtocounter{table}{-1}
  \caption{-- {\emph {continued}}}
  \begin{tabular}{lllllclllclllcll} \hline
 \multicolumn{1}{c}{\bf Methanol maser} &\multicolumn{4}{c}{\bf 6.7-GHz properties} & \multicolumn{6}{c}{\bf 12.2-GHz observations}\\
    \multicolumn{1}{c}{\bf ($l,b$)}& {\bf S$_{6.7}$} & {\bf Vp$_{6.7}$} & {\bf Vr$_{6.7}$ } & {\bf  I$_{6.7}$} & \multicolumn{2}{c}{\bf 2008 June} & \multicolumn{2}{c}{\bf 2008 December} & \multicolumn{2}{c}{\bf 2010 March}\\	
      \multicolumn{1}{c}{\bf (degrees)}  &{\bf (Jy)} && {\bf (\kmsns)}& & \multicolumn{1}{c}{\bf 5-$\sigma$} & \multicolumn{1}{c}{\bf Vr} & \multicolumn{1}{c}{\bf 5-$\sigma$} & \multicolumn{1}{c}{\bf Vr} & \multicolumn{1}{c}{\bf 5-$\sigma$} & \multicolumn{1}{c}{\bf Vr}\\  \hline \hline
G\,313.705--0.190			&	1.7	&	--41.4	&	--46.8,--41.3	&	0.8	&	$<$0.80			&	--170,170			&	$<$0.75			&	--200,140			& --\\
G\,313.767--0.863			&	14	&	--44.8	&	--45.3,--40.5	&	27	&	$<$0.80			&	--170,170			&	$<$0.75			&	--200,140			& --\\
G\,313.774--0.863			&	13	&	--56.4	&	--57.3,--52.6	&	14	&	$<$0.75			&	--170,170			&	$<$0.75			&	--200,140			& --\\
G\,313.994--0.084			&	15	&	--4.9		&	--9.2,--2.5		&	15	&	\multicolumn{6}{c}{detection}	\\
G\,314.221+0.273			&	2.8	&	--61.6	&	--61.8,--60.6	&	1.0	&	$<$0.70			&	--170,170			&	$<$0.75			&	--200,140			& --\\
G\,314.320+0.112			&	37	&	--43.4	&	--58.5,--42.7	&	28	&	$<$0.75			&	--170,170			&	$<$0.75			&	--200,140			& --\\
G\,315.803--0.575			&	8.9	&	8.2		&	7.9,8.5		&	3.0	&	$<$0.70			&	--170,170			&	$<$0.75			&	--200,140			& --\\
G\,316.359--0.362			&	98	&	3.4		&	1.1,8.4		&	77	&	\multicolumn{6}{c}{detection}	\\
G\,316.381--0.379			&	18	&	--0.7		&	--5.9,1.0		&	34	&	\multicolumn{6}{c}{detection}	\\
G\,316.412--0.308			&	11	&	--5.7		&	--6.4,--1.9		&	10	&	$<$0.75			&	--170,170			&	$<$0.80			&	--200,140		& --	\\
G\,316.484--0.310			&	0.7	&	--11.5	&	--15.6,--11.5	&	0.3	&	$<$0.75			&	--170,170			&	$<$0.75			&	--200,140		& --	\\
G\,316.640--0.087			&	96	&	--20.4	&	--25.1,--15.7	&	178	&	\multicolumn{6}{c}{detection}	\\
G\,316.811--0.057			&	51	&	--45.7	&	--48.5,--37.2	&	54	&	\multicolumn{6}{c}{detection}	\\
G\,317.029+0.361			&	0.9	&	--47.9	&	--51.1,--46.6	&	0.5	&	$<$0.75			&	--170,170			&	$<$0.75			&	--200,140			& --\\
G\,317.061+0.256			&	0.7	&	--43.8	&	--44.0,--43.5	&	0.2	&	$<$0.70			&	--170,170			&	$<$0.75			&	--200,140			& -- \\
G\,317.466--0.402*			&	51	&	--37.7	&	--50.9,--35.2	&	93	&	\multicolumn{6}{c}{detection}	\\
G\,317.701+0.110			&	23	&	--42.2	&	--47.2,--40.0	&	26	&	$<$0.70			&	--170,170			&	$<$0.75			&	--200,140		& --	\\
G\,318.043--1.404			&	6.1	&	46.3		&	44.6,46.6		&	3.0	&	$<$0.75		&	--170,170		&	$<$0.75		&	--200,140		& --\\
G\,318.050+0.087			&	11	&	--51.7	&	--56.5,--46.1	&	5.5	&	\multicolumn{6}{c}{detection}	\\
G\,318.472--0.214			&	1.1	&	--19.1	&	--20.3,--18.6	&	0.8	&	$<$0.75		&	--170,170		&	$<$0.75		&	--200,140		& --\\
G\,318.948--0.196			&	569	&	--34.6	&	--38.9,--30.8	&	526	&	\multicolumn{6}{c}{detection}	\\
G\,319.163--0.421*			&	7.0	&	--21.1	&	--22.5,--12.5	&	7.1	&	\multicolumn{6}{c}{detection}	\\
G\,319.836--0.197			&	0.2	&	--9.3		&	--13.3,--9.1	&	0.1	&	$<$0.75		&	--190,150		&	$<$0.75		&	--200,140		& --\\ 
G\,320.123--0.504			&	3.6	&	--10.1	&	--11.3,--9.4	&	2.8	&	$<$0.75		&	--190,150		&	$<$0.75		&	--200,140	& --	\\
G\,320.231--0.284*			&	54	&	--62.3	&	--71.0,--57.9	&	67	&	\multicolumn{6}{c}{detection}	\\
G\,320.244--0.562			&	1.5	&	--49.6	&	--50.0,--49.0	&	1.2	&	\multicolumn{6}{c}{detection}	\\  
G\,320.285--0.308			&	0.8	&	--69.0	&	--69.0,--61.9	&	0.5	&	$<$0.75		&	--190,150		&	$<$0.75		&	--200,140		& --\\
G\,320.414+0.109			&	2.6	&	--13.5	&	--13.6,--13.0	&	1.2	&	$<$0.75	&		--190,150	&		$<$0.75	&		--200,140	& --	\\ 
G\,320.424+0.089			&	1.9	&	--8.0		&	--9.3,--4.4		&	2.4	&	$<$0.70	&		--190,150	&		$<$0.75	&		--200,140	& --	\\
G\,320.625+0.098			&	0.5	&	--7.6		&	--8.1,--7.1		&	0.3	&	$<$0.75	&		--190,150	&		$<$0.80	&		--200,140	& --	\\
G\,320.780+0.248			&	35	&	--5.1		&	--10.3,--3.5	&	49	&	\multicolumn{6}{c}{detection}	\\
G\,321.030--0.485			&	20	&	--66.5	&	--68,--56		&	conf	&	\multicolumn{6}{c}{detection}	\\
G\,321.033--0.483			&	82	&	--61.2	&	--69,--54		&	conf	&	$<$0.80	&		--190,150	&		$<$0.80	&		--200,140	& --	\\
G\,321.148--0.529			&	9.1	&	--66.1	&	--67.4,--61.1	&	4.6	&	\multicolumn{6}{c}{detection}	\\
G\,321.704+1.168$^{sc}$		&	1.8	&	--44.3	&	--44.4,--43.7	&	0.8	&	$<$0.70	&		--190,150	&		$<$0.80	&		--200,140	& --	\\ 
G\,322.158+0.636			&	324	&--63.3		&	--66.4,--50.6	&	628	&	\multicolumn{6}{c}{detection}	\\
G\,322.705--0.331			&	2.4	&	--21.6	&	--23.5,--21.1	&	1	&	$<$0.75	&		--165,175	&		$<$0.80	&		--200,140	& --	\\
G\,323.459--0.079			&	18	&	--66.9	&	--68.3,--65.9	&	12	&	\multicolumn{6}{c}{detection}	\\
G\,323.740--0.263			&	3162	&--51.1		&	--59.7,--40.3	&	5121	&	\multicolumn{6}{c}{detection}	\\
G\,323.766--1.370			&	2.5	&	46.6		&	46.5,51.2		&	0.9	&	$<$0.75	&		--165,175	&		$<$0.75	&		--200,140		& --\\
G\,323.793--0.397			&	2.4	&	--22.6	&	--25.5,--22.4	&	0.8	&	$<$0.75	&		--165,175	&		$<$0.75	&		--200,140	& --	\\
G\,323.799+0.017			&	6.7	&	--56.1	&	--59.8,--55.8	&	3.1	&	$<$0.75	&		--165,175	&		$<$0.55	&		--200,140	& $<$0.80	& 		--155,185	\\
G\,324.716+0.342			&	10	&	--45.9	&	--50.3,--45.1	&	4.6	&	$<$0.70	&		--165,175	&		$<$0.80	&		--200,140	& --	\\
G\,324.789--0.378			&	1.2	&	11.8		&	11.7,12.1		&	0.4	&	$<$0.70	&		--165,175	&		$<$0.75	&		--200,140	& --	\\
G\,324.915+0.158			&	9.6	&	--4.9		&	--5.8,--0.6		&	25	&	\multicolumn{6}{c}{detection}	\\
G\,324.923--0.568			&	3.2	&	--78.1	&	--79.6,--77.7	&	2.2	&	\multicolumn{6}{c}{detection}	\\
G\,325.659--0.022			&	0.6	&	29.4		&	29.3,29.7		&	0.2	&	$<$0.75			&	--165,175			&	$<$0.75			&	--200,140			& --\\
G\,326.323--0.393			&	3.2	&	--69.3	&	--77.1,--67.6	&	2.7	&	$<$0.75			&	--165,175			&	$<$0.75			&	--200,140		& --	\\
G\,326.448--0.748			&	4.5	&	--68.5	&	--73.5,--57.5	&	12	&	\multicolumn{6}{c}{detection}	\\
G\,326.475+0.703			&	129	&	--38.5	&	--50.6,--36.7	&	135	&	\multicolumn{6}{c}{detection}	\\ 
G\,326.476+0.695			&	3.3	&	--43.5	&	--43.8,--43.1	&	1.3	&	--				&					&	$<$0.80			&	--200,140			& $<$0.80	&--155,185\\
G\,326.608+0.799			&	1.3	&	--44.9	&	--45.1,--44.3	&	0.5	&	$<$0.75			&	--165,175			&	$<$0.80			&	--200,140		& --	\\
G\,326.641+0.611*			&	31	&	--42.5	&	--45.3,--34.9	&	25	&	\multicolumn{6}{c}{detection}	\\
G\,326.662+0.520			&	16	&	--38.6	&	--42.7,--38.3	&	21	&	$<$0.75			&	--165,175			&	$<$0.80			&	--200,140		& --	\\
G\,326.859--0.677			&	15	&	--58.1	&	--62.4,--57.1	&	12	&	\multicolumn{6}{c}{detection}	\\
G\,326.986--0.031			&	2.3	&	--60.7	&	--60.8,--56.6	&	1.4	&	$<$0.70			&	--165,175			&	$<$0.75			&	--200,140	 & --		\\
G\,327.120+0.511			&	55	&	--87.1	&	--92.4,--82.9	&	48	&	\multicolumn{6}{c}{detection}	\\
G\,327.282--0.469			&	5.4	&	0.1		&	--1.3,1.8		&	4.5	&	\multicolumn{6}{c}{detection}	\\
G\,327.291--0.578			&	2.5	&	--37.0	&	--45.4,--36.8	&	2.4	&	\multicolumn{6}{c}{detection}	\\
G\,327.392+0.199			&	8.4	&	--84.1	&	--86.1,--79.5	&	10	&	\multicolumn{6}{c}{detection}	\\
G\,327.395+0.197			&	9.2	&	--88.9	&	--89.7,--88.3	&	3.5	&	\multicolumn{6}{c}{detection}	\\
G\,327.402+0.445			&	81	&	--82.6	&	--84.4,--69.9	&	135	&	\multicolumn{6}{c}{detection}	\\
G\,327.566--0.850			&	12	&	--28.8	&	--30.4,--21.6	&	9.7	&	$<$0.75			&	--165,175			&	$<$0.80			&	--200,140	&	 --	\\
\end{tabular}\label{tab:6MMB}
\end{table*}

\begin{table*}\addtocounter{table}{-1}
  \caption{-- {\emph {continued}}}
  \begin{tabular}{lllllclllclllcll} \hline
 \multicolumn{1}{c}{\bf Methanol maser} &\multicolumn{4}{c}{\bf 6.7-GHz properties} & \multicolumn{6}{c}{\bf 12.2-GHz observations}\\
    \multicolumn{1}{c}{\bf ($l,b$)}& {\bf S$_{6.7}$} & {\bf Vp$_{6.7}$} & {\bf Vr$_{6.7}$ } & {\bf  I$_{6.7}$} & \multicolumn{2}{c}{\bf 2008 June} & \multicolumn{2}{c}{\bf 2008 December} & \multicolumn{2}{c}{\bf 2010 March}\\	
      \multicolumn{1}{c}{\bf (degrees)}  &{\bf (Jy)} &&{\bf (\kmsns)}& & \multicolumn{1}{c}{\bf 5-$\sigma$} & \multicolumn{1}{c}{\bf Vr} & \multicolumn{1}{c}{\bf 5-$\sigma$} & \multicolumn{1}{c}{\bf Vr} & \multicolumn{1}{c}{\bf 5-$\sigma$} & \multicolumn{1}{c}{\bf Vr}\\  \hline \hline
G\,327.590--0.094			&	4.0	&	--86.3	&	--86.6,--85.7	&	1.7	&	$<$0.70			&	--165,175			&	$<$0.80			&	--200,140	&	--	\\  
      G\,327.618--0.111		&	2.1	&	--98.1	&	--98.8,--94.3	&	1.9	&	$<$0.70			&	--165,175			&	$<$0.80			&	--200,140	&	 --	\\
G\,327.710--0.394			&	4.2	&	--77.6	&	--78.6,--72.2	&	3.2	&	$<$0.70			&	--165,175			&	$<$0.80			&	--200,140	&	 --	\\
G\,327.808--0.634			&	2.7	&	--42.3	&	--42.6,--41.9	&	1.2	&	$<$0.70			&	--165,175			&	$<$0.80			&	--200,140	&	 --	\\
G\,327.863+0.098			&	1.6	&	--46.2	&	--47.4,--45.9	&	0.9	&	$<$0.70			&	--165,175			&	$<$0.80			&	--200,140	&	 --	\\
G\,327.945--0.115			&	7.8	&	--51.6	&	--52.0,--50.8	&	3.6	&	\multicolumn{6}{c}{detection}	\\
G\,328.140--0.432			&	16	&	--39.6	&	--40.2,--36.3	&	20	&	$<$0.75			&	--165,175			&	$<$0.75			&	--200,140	& --\\
G\,328.164+0.587			&	1.6	&	--91.9	&	--92.1,--91.3	&	0.5	&	$<$0.75		&	--165,175	&		$<$0.80		&	--200,140	& --	\\
G\,328.237--0.547			&	1340	&	--44.7	&	--46.2,--31.7	&	808	&	\multicolumn{6}{c}{detection}	\\
G\,328.254--0.532			&	361	&	--37.4	&	--50.4,--35.5	&	236	&	\multicolumn{6}{c}{detection}	\\
G\,328.385+0.131			&	1.6	&	28.9		&	28.7,29.2		&	0.6	&	$<$0.75		&	--165,175	&		$<$0.80		&	--200,140		& --\\
G\,328.808+0.633*			&	296	&	--43.8	&	--47,--42		&	conf	&	\multicolumn{6}{c}{detection}	\\
G\,328.809+0.633*			&	346	&	--44.2	&	--45,--43		&	conf	&	\multicolumn{6}{c}{detection}	\\
G\,328.819+1.704			&	1.5	&	--86.3	&	--88.7,--71.2	&	1.4	&	$<$0.75		&	--165,175	&		$<$0.80		&	--200,140	& --	\\
G\,328.940+0.558			& 	0.8	&	--98.8	&	--98.9,--87.7	&	0.9	&	\multicolumn{6}{c}{detection}	\\
G\,328.942+0.565			&	2.1	&	--90.9	&	--91.3,--89.1	&	1.9	&	\multicolumn{6}{c}{detection}	\\ 
G\,329.029--0.205			&	113	&	--36.9	&	--42.0,--33.5	&	210	&	\multicolumn{6}{c}{detection}	\\
G\,329.031--0.198			&	32	&	--46.0	&	--48.1,--42.0	&	36	&	$<$0.75		&	--165,175	&			--	&			& --\\
G\,329.066--0.308			&	14	&	--43.9	&	--48.6,--40.2	&	18	&	\multicolumn{6}{c}{detection}	\\
G\,329.183--0.314			&	5.1	&	--55.6	&	--59.9,--50.2	&	8.7	&	\multicolumn{6}{c}{detection}	\\
G\,329.272+0.115			&	1.3	&	--72.0	&	--72.1,--71.9	&	0.3	&	\multicolumn{6}{c}{detection}	\\
G\,329.339+0.148			&	24	&	--106.4	&	--107.5,--105.5	&	13	&	\multicolumn{6}{c}{detection}	\\
G\,329.341--0.644			&	3.0	&	--81.3	&	--81.6,--81.1	&	1.0	&	$<$0.70		&	--165,175	&			$<$0.80	&			--200,140		& --	\\
G\,329.405--0.459			&	51	&	--70.4	&	--73.3,--63.2	&	conf	&	\multicolumn{6}{c}{detection}	\\
G\,329.407--0.459			&	96	&	--66.7	&	--68.0,--66.0	&	conf	&	\multicolumn{6}{c}{detection}	\\
G\,329.469+0.503			&	18	&	--72.0	&	--73.7,--58.8	&	25	&	$<$0.70	&			--165,175	&			$<$0.80	&			--200,140			& --\\
G\,329.526+0.216			&	1.8	&	--92.8	&	--93.2,--92.5	&	0.7	&	$<$0.75	&			--165,175	&			$<$0.80	&			--200,140		& --	\\
G\,329.556+0.181			&	1.5	&	--106.5	&	--109.1,--92.6	&	1.8	&	$<$0.75	&			--165,175	&			$<$0.80	&			--200,140			& --\\
G\,329.610+0.114			&	44	&	--60.1	&	--68.9,--58.6	&	32	&	\multicolumn{6}{c}{detection}	\\
G\,329.622+0.138			&	2.0	&	--85.0		&	--85.7,--83.1	&	1.0	&	\multicolumn{6}{c}{detection}	\\
G\,329.719+1.164			&	7.7	&	--75.9	&	--82.5,--72.8	&	9.4	&	\multicolumn{6}{c}{detection}	\\

\end{tabular}\label{tab:6MMB}
\end{table*}

\begin{table*}
\caption{Characteristics of the 12.2-GHz methanol maser emission that we detect. Column 1 gives the Galactic longitude and latitude of each source and is used as an identifier; and columns 2 and 3 give the equatorial coordinates for each of the sources and have been derived from interferometric observations of the 6.7-GHz methanol masers detected in the MMB survey \citep{Green12}. References for previously detected 12.2-GHz sources follow the source name and are as follows; 1: \citet{Breen10a}; 2: \citet{Caswell95b}; 3: ~\citet{Gay}; 4: \citet{Koo88}; 5: \citet{Catarzi93}; 6: \citet{Kemball88}; 7: \citet{Cas93}; 8: \citet{Norris1987}; 9: \citet{MacLeod93}; and are presented as superscripts after the source name. Column 4 gives the epoch of the observed data (sometimes indicating that the presented data is the average of multiple epochs) presented in columns 5 - 8 which give the peak flux density (Jy) or 5-$\sigma$ detection limit, velocity of the 12.2-GHz peak (\kmsns), velocity range (\kmsns) and integrated flux density (Jy \kmsns), respectively. The presence of a `--' in any of the columns showing source flux density indicates that no observation was made at the given epoch. The presence of `conf' in the columns showing the integrated flux density indicates that a value for this characteristic could not be extracted do to confusion from nearby sources. Detection limits that are followed by `$^{*}$' indicate that while no detection was made at the given epoch, emission is detected in the average spectrum. Flux densities derived from averaged spectrums are followed by a `$^{\alpha}$', and for the one source that we interpret to be thermal (G\,208.996--19.386), a `$^{\gamma}$'.} 
  \begin{tabular}{lllllccclclllclllcll} \hline
\multicolumn{1}{c}{\bf Methanol maser} &\multicolumn{1}{c}{\bf RA} & \multicolumn{1}{c}{\bf Dec}  &{\bf Epoch}&  {\bf S$_{12.2}$} & {\bf Vp$_{12.2}$} & {\bf Vr$_{12.2}$} & \multicolumn{1}{c}{\bf I$_{12.2}$} \\
    \multicolumn{1}{c}{\bf ($l,b$)}&  \multicolumn{1}{c}{\bf (J2000)} & \multicolumn{1}{c}{\bf (J2000)}   & & {\bf (Jy)} & {\bf (\kmsns)} & {\bf  (\kmsns)} & \multicolumn{1}{c}{\bf (Jy \kmsns)}\\
    \multicolumn{1}{c}{\bf (degrees)}  & \multicolumn{1}{c}{\bf (h m s)}&\multicolumn{1}{c}{\bf ($^{o}$ $'$ $``$)}\\  \hline \hline
G\,188.946+0.886$^{2,4,5,6,7}$		&	06 08 53.32	&	+21 38 29.1	&	2008 Jun	&	225	&	10.9	& 9.5,12.0	&150	\\
						&				&				&	2008 Dec	&	230	&	10.9	&	9.4,12.0	&	152\\
G\,192.600--0.048$^{1}$		&	06 12 53.99	&	+17 59 23.7	&	2008 Jun	& 0.5		&	3.6	&	3.4,3.7	&	0.1	\\
						&				&				&	2008 Dec	&	0.6	&	3.6	&	3.3,3.8	&	0.2	\\ 
G\,208.996--19.386			&	05 35 14.50	&	--05 22 45.0	&	2008 Dec	&$<$0.75*	&		&	--150,100\\
						&				&				&	2010 Mar	&$<$0.90*	&		&	--150,100		\\
						&				&				&	Avg all	&	0.3$^{\alpha\gamma}$	&	8.4	&	7.6,9.0	&	0.3	\\
G\,213.705--12.597$^{1,2,9}$	&	06 07 47.85	&	--06 22 55.2	&	2008 Jun	&	14	&	12.7	&	12.3,13.3	&	6.0	\\
						&				&				&	2010 Mar	&	11	&	12.8	&	12.3,13.3	&	4.5	\\
G\,232.620+0.996$^{2}$		&	07 32 09.79	&	--16 58 12.4	&	2008 Jun	&	27	&	22.9	&	22.0,23.3	&	8.6	\\
						&				&				&	2010 Mar	&	15	& 	22.8	&	22.1,23.3	&	4.9	\\
G\,263.250+0.514			&	08 48 47.84	&	 --42 54 28.3	&	2008 Jun	&	0.4	&	12.4	&	12.3,12.5	&	0.1	\\
						&				&				&	2008 Dec	&	0.7	&	12.5	&	12.2,12.5	&	0.2	\\
						&				&				&	2010 Mar	& 	0.4	&	12.5	&	12.3,12.5	&	0.1\\
G\,264.289+1.469			&	08 56 26.80	&	 --43 05 42.1	&	2008 Jun	&	0.3	&	8.7	&	8.6,8.8	&	0.1\\
						&				&				&	2008 Dec	&	$<$0.80	&		&	--205,135	&		\\
						&				&				&	2010 Mar	& 	0.8	&	6.4	&	6.3,8.2	&	0.1\\
G\,269.456--1.467$^{1}$		&	09 03 14.78	&	 --48 55 11.2	&	2008 Jun	&	1.4	&	56	&	55.8,56.4	&	0.4	\\
						&				&				&	2010 Mar	&	$<$0.73	&	&	--200,140\\
G\,285.337--0.002			&	10 32 09.62	&	 --58 02 04.6	&	2008 Jun	&	1.4	&	0.6	&	0.4,1.1	&	0.5	\\
						&				&				&	2010 Mar	&	1.8	&	0.6	&	0.4,0.8	&	0.5\\
G\,286.383--1.834			&	10 31 55.12	&	 --60 08 38.6	&	2008 Jun	&	4.3	&	9.6	&	9.2,9.9	&	1.7	\\
						&				&				&	2010 Mar	&	4.7	&	9.6	&	9.2,9.8	&	1.8\\
G\,287.371+0.644$^{1,3}$	&	10 48 04.44	&	 --58 27 01.0	&	2008 Jun	&	49	&	--2.0	&	--2.6,--0.9	&	20	\\
						&				&				&	2008 Dec	&	52	&	--2.0	&	--2.6,--0.9	&	22	\\
						&				&				&	2010 Mar	&	51	&	--2.0	& --2.6,--1.0	&20 	\\
G\,290.411--2.915$^{1}$		&	10 57 33.89	&	 --62 59 03.5	&	2008 Jun	&	1.6	&	--16.1	&	--16.6,--15.9	&	0.4	\\
						&				&				&	2010 Mar	& 	1.7	&	--16.1	&	--16.2,--16.0	&	0.4\\
G\,291.274--0.709$^{1,2,7,8}$	&	11 11 53.35	&	 --61 18 23.7	&	2008 Jun	&	3.0	&	--29.1	&	--30.1,--28.7	&	2.3	\\
						&				&				&	2010 Mar	&	1.7	&--29.2	&	--30.1,--28.9	& 1.4\\
G\,291.579--0.431			&	11 15 05.76	&	 --61 09 40.8	&	2008 Jun	&	$<$0.60	&		&	--200,140	&		\\
						&				&				&	2008 Dec	&	$<$0.55	&		&	--200,140	&	\\
						&				&				&	2010 Mar	&	 0.5	&	15.3	&	15.2,15.4	& 0.1	\\ 
G\,293.723--1.742			&	11 28 32.97	&	 --63 07 18.6	&	2008 Jun	&	0.6	&	24.3	&	24.1,24.4	&	0.1	\\
						&				&				&	2008 Dec	&	$<$0.80	&		&	--200,140	&		\\
						&				&				&	2010 Mar	& $<$0.55		&	&	--180,160\\ 
G\,293.942--0.874$^{1}$		&	11 32 42.09	&	 --62 21 47.5	&	2008 Jun	&	2.2	&	41.0	&	40.7,41.3	&	0.6	\\
						&				&				&	2010 Mar	& 	1.4	&	41.0	&	40.9,41.2	&	0.3\\
G\,298.632--0.362			&	12 13 31.63	&	 --62 55 01.0	&	2008 Jun	&	0.3	&	37.3	&	36.9,37.3	&	0.1	\\
						&				&				&	2008 Dec	&	$<$0.85	&		&	--200,140	&		\\
						&				&				&	2010 Mar	& 0.5 &	37.2	& 37.2,37.3	& 0.1\\ 
G\,299.772--0.005			&	12 23 48.97	&	 --62 42 25.3	&	2008 Jun	&	0.4	&	--7.0		&--7.3,--7.0	&	0.1	\\
						&				&				&	2008 Dec	&	$<$0.55	&		&	--200,140	&	\\
						&				&				&	2010 Mar	& $<$0.55	& 	&	--180,160	\\ 
G\,300.969+1.148$^{2}$		&	12 34 53.29	&	 --61 39 40.0	&	2008 Jun	&	0.5	&	--36.8	&	--37.0,--36.8	&	0.1	\\
						&				&				&	2010 Mar & 0.7		&	--36.9	&	--37.0,--36.5	&	0.3	\\
G\,303.846--0.363			&	12 59 33.37	&	 --63 13 14.7	&	2008 Jun	&	2.0	&	24.8	&	24.6,26.9	&	0.6	\\
						&				&				&	2010 Mar	& 3.8	&	24.7	&	24.5,27.2	& 1.4\\ 
\end{tabular}\label{tab:12MMB}

\end{table*}
\clearpage

\begin{table*}\addtocounter{table}{-1}
  \caption{-- {\emph {continued}}}
  \begin{tabular}{lllllccclclllclllcll} \hline
\multicolumn{1}{c}{\bf Methanol maser} &\multicolumn{1}{c}{\bf RA} & \multicolumn{1}{c}{\bf Dec}  &{\bf Epoch}&  {\bf S$_{12.2}$} & {\bf Vp$_{12.2}$} & {\bf Vr$_{12.2}$} & \multicolumn{1}{c}{\bf I$_{12.2}$} \\
    \multicolumn{1}{c}{\bf ($l,b$)}&  \multicolumn{1}{c}{\bf (J2000)} & \multicolumn{1}{c}{\bf (J2000)}   & & {\bf (Jy)} & {\bf (\kmsns)} & {\bf  (\kmsns)} & \multicolumn{1}{c}{\bf (Jy \kmsns)}\\
    \multicolumn{1}{c}{\bf (degrees)}  & \multicolumn{1}{c}{\bf (h m s)}&\multicolumn{1}{c}{\bf ($^{o}$ $'$ $``$)}\\  \hline \hline
G\,305.199+0.005$^{1}$		&	13 11 17.20	&	 --62 46 46.0	&	2008 Jun	&	0.6	&	--42.7	&	--42.8,--42.6	&	0.1	\\
						&				&				&	2010 Mar	& 0.6		&	--42.7	& --43.3,--41.6	& 0.3\\ 
G\,305.200+0.019$^{1,2}$	&	13 11 16.93	&	 --62 45 55.1	&	2008 Jun	&	11	&	--31.9	&	--33.2,--31.6	&	8.8	\\
						&				&				&	2010 Mar	& 8.3	& --32.1	&	--33.3,--31.6	& 8.0\\
G\,305.202+0.208$^{1,2,7,8,9}$	&	13 11 10.49	&	 --62 34 38.8	&	2008 Jun &	3.5	&	--43.9	&	--45.0,--43.6	&	1.8	\\
							&				&				&	2008 Dec	&	1.0	&	--43.9	&	--45.2,--43.7	&	0.6	\\
G\,305.208+0.206$^{1,2,7,8,9}$	&	13 11 13.71	&	 --62 34 41.4	&	2008 Jun	&	1.8	&	--36.5	&	--40.4,--35.9	&	1.7	\\
							&				&				&	2008 Dec	&	2.0	&	--36.5	&	--40.4,--35.9	&	1.9	\\
G\,305.366+0.184$^{1,2}$	&	13 12 36.74	&	 --62 35 14.7	&	2008 Jun	&	3.3	&	--34.3	&	--34.8,--33.6	&	1.6	\\
						&				&				&	2010 Mar	& 2.3	&	--34.3	& --34.5,--33.7	& 1.1		\\
G\,305.563+0.013$^{1}$		&	13 14 26.90	&	 --62 44 29.4	&	2008 Jun	&1.5	&	--37.2	&	--37.4,--32.5	&	1.3	\\
						&				&				&	2010 Mar	& 2.2	&	--37.0	& --41.9,--35.6	&	2.1	\\
G\,305.940--0.164			&	13 17 53.05	&	 --62 52 50.5	&	2008 Jun	&	0.7	&	--51.0	&	--51.1,--50.9	&	0.1	\\
						&				&				&	2008 Dec	&	$<$0.75	&		&	--200,140	&		\\
						&				&				&	2010 Mar	&	0.9	&	--51.1	&	--51.3,--51.0	& 0.2\\
G\,308.651--0.507			&	13 41 50.19	&	 --62 49 05.2	&	2008 Jun	&	1.0	&	3.0	&	2.7,3.2	&	0.4	\\
						&				&				& 	2010 Mar	&	0.7	&	2.8	&	2.7,3.1	&	0.2\\
G\,308.754+0.549			&	13 40 57.60	&	 --61 45 43.4	&	2008 Jun	&	0.5	&	--51.2	&	--51.2,--43.4	&	0.1	\\
						&				&				&	2008 Dec	&	$<$0.80	&		&	--200,140	&	&	\\
						&				&				&	2010 Mar	&	0.5	&	--51.5	&	--51.3,--50.3	&	0.1\\
G\,308.918+0.123$^{2,7,8}$	&	13 43 01.85	&	 --62 08 52.2	&	2008 Dec	&	8.2	&	--53.3	&	--54.7,--53.0	&	4.6	\\
						&				&				&	2010 Mar	& 8.7		&	--53.3	&--54.6,--53.0	&	4.8	\\
G\,309.901+0.231			&	13 51 01.05	&	 --61 49 56.0	&	2008 Dec	&	5.6	&	--54.6	&	--55.6,--53.9	&	2.1	\\
						&				&				&	2010 Mar	&	4.1	&	--54.5	& --54.9,--53.9	&	1.8\\
G\,309.921+0.479$^{1,2,7,8}$	&	13 50 41.78	&	 --61 35 10.2	&	2008 Jun	&	179	&	--59.8	&	--61.3,--57.7	&	108	\\
						&				&				&	2010 Mar	& 	159	&	--59.7	&	--61.7,--57.7	&	95\\
G\,310.144+0.760$^{3}$		&	13 51 58.43	&	 --61 15 41.3	&	2008 Jun	&	50	&	--55.7	&	--57.8,--54.6	&	47\\
						&				&				&	2010 Mar	&	49	&	--56.0	&	--57.7,--54.7	&	52	\\
G\,311.551--0.055			&	14 05 07.06	&	 --61 41 15.9	&	2008 Jun	&	$<$0.80*	&		&	--170,170	&		\\
						&				&				&	2008 Dec	&	$<$0.75*	&		&	--200,140	&		\\
						&				&				&	Avg all	&	0.4$^{\alpha}$		&	--56.3	&	--56.4,--56.3	&	0.1	\\ 
G\,311.628+0.266			&	14 04 59.20	&	 --61 21 29.4	&	2008 Jun	&	0.6	&	--57.5	&	--57.6,--57.4	&	0.2	\\
						&				&				&	2008 Dec	&	0.9	&	--57.5	&	--57.9,--57.3	&	0.2	\\
						&				&				&	2010 Mar	&	1.0	&	--57.5	& --58.0,--56.5	& 0.4\\
G\,312.071+0.082			&	14 08 58.20	&	 --61 24 23.8	&	2008 Jun	&	12	&	--34.8	&	--35.2,--32.4	&	10	\\
						&				&				&	2010 Mar	& 	11	&	--34.8	&	--35.2,--32.3	&	8.9\\
G\,313.469+0.190$^{2}$		&	14 19 40.94	&	 --60 51 47.3	&	2008 Dec 	&	4.9	&	--9.3	&	--12.4,--8.0	&	4.3	\\
						&				&				&	2010 Mar	& 2.8		&	--9.3	&	--12.4,--7.9	&	2.8\\
G\,313.577+0.325$^{3}$		&	14 20 08.58	&	 --60 42 00.8	&	2008 Jun	&	$<$0.75	&		&	--170,170	&		\\
						&				&				&	2008 Dec	&	0.7	&	--47.8	&	--47.9,--47.6	&	0.2	\\
						&				&				&	2010 Mar	&	$<$0.80	& &--165,175\\ 
G\,313.994--0.084			&	14 24 30.78	&	 --60 56 28.3	&	2008 Jun	&	1.6	&	--4.8	&	--8.1,--4.4	&	1.0	\\
						&				&				&	2010 Mar	&	 1.5	& --4.8	& --7.9,--4.6	& 0.7\\
G\,316.359--0.362$^{2,3}$	&	14 43 11.20	&	 --60 17 13.3	&	2008 Dec 	&	9.9	&	3.3	&	1.3,7.3	&	6.8	\\
						&				&				&	2010 Mar	& 9.6	& 3.4	& 1.3,7.2	&	5.5\\
G\,316.381--0.379$^{2}$		&	14 43 24.21	&	 --60 17 37.4	&	2008 Jun	&	1.7	&	--0.6	&	--4.8,0.7	&	2.2	\\
G\,316.640--0.087$^{2,3,7,8}$	&	14 44 18.45	&	 --59 55 11.5	&	2008 Dec 	&	19	&	--20.0	&	--22.6,--16.5	&	35	\\
						&				&				&	2010 Mar	& 	16	& --22.1	& --22.7,--16.6& 26\\
G\,316.811--0.057$^{2}$		&	14 45 26.43	&	 --59 49 16.3	&	2008 Dec	&	18	&	--46.5	&	--47.5,--45.8	&	13	\\
						&				&				&	2010 Mar	& 	12 	& --46.6	& --48.1,--45.7	& 9.4\\
G\,317.466--0.402			&	14 51 19.69	&	 --59 50 50.7	&	2008 Jun	&	5.6	&	--37.6	&	--41.5,--35.6	&	8.5	\\
						&				&				&	2010 Mar	& 4.5		& --38.1	& --41.5,--35.5	& 6.4\\
G\,318.050+0.087$^{2,3}$	&	14 53 42.67	&	 --59 08 52.4	&	2008 Dec	&	2.3	&	--51.7	&	--52.0,--51.4	&	0.8\\
						&				&				&	2010 Mar	& 1.1	& --51.7	& --51.9,--51.5	& 0.4\\
G\,318.948--0.196$^{1,2,6,7}$	&	15 00 55.40	&	 --58 58 52.1	&	2008 Jun	&	139	&	--34.5	&	--35.6,--32.1	&	91	\\
						&				&				&	2010 Mar	& 	121	&	--34.5	&	--35.6,--32.0	& 79	\\
G\,319.163--0.421			&	15 03 13.74	&	 --59 04 30.5	&	2008 Dec	&	6.9	&	--22.1	&	--22.5,--20.9	&	4.7\\
						&				&				&	2010 Mar	& 6.0		& -22.1	& --22.5,--20.7	& 4.4	\\
G\,320.231--0.284			&	15 09 51.94	&	 --58 25 38.5	&	2008 Jun	&	0.5	&	--66.4	&	--66.5,--66.3	&	0.1	\\
						&				&				&	2008 Dec	&	0.3	&	--63.2	&	--63.4,--63.2	&	0.1	\\
						&				&				&	2010 Mar	& $<$0.55	&	& 	--165,175 \\ 

\end{tabular}\label{tab:12MMB}

\end{table*}
\clearpage

\begin{table*}\addtocounter{table}{-1}
  \caption{-- {\emph {continued}}}
  \begin{tabular}{lllllccclclllclllcll} \hline
\multicolumn{1}{c}{\bf Methanol maser} &\multicolumn{1}{c}{\bf RA} & \multicolumn{1}{c}{\bf Dec}  &{\bf Epoch}&  {\bf S$_{12.2}$} & {\bf Vp$_{12.2}$} & {\bf Vr$_{12.2}$} & \multicolumn{1}{c}{\bf I$_{12.2}$} \\
    \multicolumn{1}{c}{\bf ($l,b$)}&  \multicolumn{1}{c}{\bf (J2000)} & \multicolumn{1}{c}{\bf (J2000)}   & & {\bf (Jy)} & {\bf (\kmsns)} & {\bf  (\kmsns)} & \multicolumn{1}{c}{\bf (Jy \kmsns)}\\
    \multicolumn{1}{c}{\bf (degrees)}  & \multicolumn{1}{c}{\bf (h m s)}&\multicolumn{1}{c}{\bf ($^{o}$ $'$ $``$)}\\  \hline \hline
G\,320.244--0.562			&	15 11 01.61	&	 --58 39 37.7	&	2008 Jun	&	0.6	&	--49.8	&	--49.9,--49.1	&	0.2	\\
						&				&				&	2008 Dec	&	$<$0.55	&		&	--200,140	&		\\
						&				&				&	2010 Mar	& 0.4	& --49.2 	& --49.2,--49.1	& 0.1\\  
G\,320.780+0.248			&	15 11 23.48	&	 --57 41 25.1	&	2008 Dec	&	12	&	--5.2	&	--6.1,--4.2	&	13	\\
						&				&				&	2010 Mar	& 8.2		& --5.2	& --5.9,--4.3	& 6.8	\\
G\,321.030--0.485$^{3}$		&	15 15 51.79	&	 --58 11 18.0	&	2008 Jun	&	16	&	--66.6	&	--67.2,--65.4	&	9.0	\\
						&				&				&	2010 Mar	& 1.2	& --66.3	& --66.9,--66.2	&	0.6	\\ 
G\,321.148--0.529$^{2}$		&	15 16 48.39	&	 --58 09 50.2	&	2008 Jun	&	1.4	&	--66.1	&	--66.3,--65.8	&	0.4	\\
						&				&				&	2010 Mar	& 0.7	& --66.2 & --66.3,--66.0	& 0.2	\\
G\,322.158+0.636$^{2,7,9}$	&	15 18 34.64	&	 --56 38 25.3	&	2008 Dec	&13	&	--63.4	&	--64.1,--52.9	&	15			\\
G\,323.459--0.079$^{2,6,7}$	&	15 29 19.33	&	 --56 31 22.8	&	2008 Dec	&	11	&	--66.7	&	--67.4,--66.5	&	3.8       \\
						&				&				&	2010 Mar	& 11	& 	--66.7	& --67.4,--66.5	& 3.5	\\
G\,323.740--0.263$^{1,2,6,7,8}$&	15 31 45.45	&	 --56 30 50.1	&	2008 Jun	&	396	&	--48.7	&	--52.4,--48.1	&	684	\\
G\,324.915+0.158			&	15 36 51.17	&	 --55 29 22.9	&	2008 Jun	&	1.2	&	--3.7	&	--5.0,--1.5	&	1.5	\\
G\,324.923--0.568			&	15 39 57.64	&	 --56 04 08.3	&	2008 Jun	&	1.0	&	--78.2	&	--79.7,--78.1	&	0.3	\\
G\,326.448--0.748			&	15 49 18.63	&	 --55 16 51.6	&	2008 Jun	&	1.2	&	--60.9	&	--71,--57.7	&	0.8		\\
G\,326.475+0.703			&	15 43 16.64	&	 --54 07 14.6	&	2008 Dec	&	17	&	--38.5	&	--49.5,--37.6	&	9.1	\\ 
G\,326.641+0.611			&	15 44 33.33	&	 --54 05 31.5	&	2008 Jun	&	0.7	&	--39.9	&	--40.1,--39.5	&	0.3	\\
						&				&				&	2010 Mar	& $<$0.70		& --155,185\\ 
G\,326.859--0.677			&	15 51 14.19	&	 --54 58 04.8	&	2008 Jun	&	2.1	&	--58.0	&	--58.3,--57.9	&	0.7	\\
G\,327.120+0.511$^{2}$		&	15 47 32.73	&	 --53 52 38.4	&	2008 Dec	&	2.9	&	--89.6	&	--90.1,--89.2	&	1.2	\\
G\,327.282--0.469			&	15 52 36.03	&	 --54 32 24.0	&	2008 Jun	&	1.1	&	0.2	&	--0.3,0.3	&	0.6	\\
						&				&				&	2008 Dec	&	1.5	&	--0.1	&	--0.2,0.2	&	0.5 	\\
G\,327.291--0.578$^{2}$		&	15 53 07.65	&	 --54 37 06.1	&	2008 Dec	&$<$0.80*	&		&	--200,140	&	\\
						&				&				&	2010 Mar	& $<$0.80*	& & --155,185	\\
						&				&				&	Avg all	&	0.4$^{\alpha}$	& --40.4	&	--40.5,--40.4	&	0.1\\ 
G\,327.392+0.199			&	15 50 18.48	&	 --53 57 06.3	&	2008 Jun	&	0.7	&	--83.2	&	--84.6,--81.9	&	0.5	\\
						&				&				&	2008 Dec	&	0.8	&	--83.3	&	--83.5,--81.9	&	0.7	\\
G\,327.395+0.197			&	15 50 20.06	&	 --53 57 07.5	&	2008 Jun	&$<$0.80*	&		&	--165,175	&		\\
						&				&				&	2008 Dec	&	$<$0.80*	&		&	--200,140	&\\
						&				&				&	Avg all	&	0.3$^{\alpha}$	&	--89.6	&	--90.1,--89.6	&	0.1\\ 
G\,327.402+0.445$^{2,7}$	&	15 49 19.50	&	 --53 45 13.9	&	2008 Dec	&	55	&	--82.7	&	--84.1,--75.2	&	35	\\
G\,327.945--0.115			&	15 54 33.91	&	 --53 50 44.3	&	2008 Jun	&	3.3	&	--51.7	&	--51.9,--51.2	&	1.3\\
G\,328.237--0.547$^{2,7,8}$	&	15 57 58.28	&	 --53 59 22.7	&	2008 Dec 	&	26	&	--44.7	&	--45.2,--44.1	&	9.8	\\
G\,328.254--0.532$^{2,7,8}$	&	15 57 59.75	&	 --53 58 00.4	&	2008 Dec	&	6.4	&	--36.8	&	--37.7,--36.4	&	2.8 	\\

G\,328.808+0.633$^{2,7,8}$	&	15 55 48.45	&	 --52 43 06.6	&	2008 Jun	&	22	&	--46.4	&	--46.9,--43.3	&	10	\\
G\,328.809+0.633$^{2,7,8}$	&	15 55 48.70	&	 --52 43 05.5	&	2008 Jun	&	5.4	&	--43.5	&	--44.6,--43.3	&	1.9		\\
G\,328.940+0.558			&	15 56 47.43	&	--52 41 28.2	& 	2008 Jun	&	0.5	&	--98.8	&	--98.9,--89.0	&	0.1	\\
						&				&				&	2008 Dec	&	0.7	&	--98.8	&	--98.9,--98.7	&	0.2 \\
						&				&				&	2010 Mar	& 0.4		&	--98.8	&	--98.8,--98.7	&	0.1\\ 
G\,328.942+0.565			&	15 56 46.17	&	 --52 41 04.8	&	2008 Jun	&$<$0.75	& 	& --165,175	& 	\\
						&				&				&	2008 Dec	& 0.6	& --90.5 & --90.5,--90.4	& 0.1 \\
						&				&				&	2010 Mar	& $<$0.80 & & --150,190&		\\  
G\,329.029--0.205$^{2,7,8}$	&	16 00 31.80	&	 --53 12 49.6	&	2008 Jun	&16	&	--37.3	&	--40.3,--36.8	&	7.8	\\
G\,329.066--0.308			&	16 01 09.93	&	 --53 16 02.6	&	2008 Jun	&	0.6	&	--43.5	&	--43.8,--43.5	&	0.2	\\
						&				&				&	2008 Dec	&	0.9	&	--43.5	&	--43.7,--43.4	&	0.2	\\
G\,329.183--0.314$^{2}$		&	16 01 47.01	&	 --53 11 43.3	&	2008 Jun	&	1.3	&	--58.2	&	--59.3,--58.1	&	0.3	\\
G\,329.272+0.115			&	16 00 21.77	&	 --52 48 48.1	&	2008 Jun	&	$<$0.70*	&		&	--165,175	&		\\
						&				&				&	2008 Dec	&	$<$0.80*	&		&	--200,140	&		\\
						&				&				&	Avg all	&	0.4$^{\alpha}$		& --72.0	&	--72.1,--71.9	&	0.1\\	
G\,329.339+0.148			&	16 00 33.13	&	 --52 44 39.8	&	2008 Jun	&	3.2	&	--106.3	&	--107.3,--106	&	1.3	\\
G\,329.405--0.459$^{2,9}$	&	16 03 32.16	&	 --53 09 30.5	&	2008 Jun	&	0.7	&	--71.6	&	--71.7,--63.6	&	0.3	\\
						&				&				&	2008 Dec	&	0.7	&	--71.7	&	--71.8,--70.1	&	0.2	\\
						&				&				&	2010 Mar	& 	0.5	& --72.2	& --72.3,--71.6 & 0.2	\\
G\,329.407--0.459$^{2,9}$	&	16 03 32.65	&	 --53 09 26.9	&	2008 Jun	&	0.5	&	--68.3	&	--68.3,--65.9	&	0.2	\\
						&				&				&	2008 Dec	&	0.4	&	--68.3	&	--68.3,--68.0	&	0.1	\\
						&				&				&	2010 Mar	& 1.1		& --67.9&--68.3,--66.3	& 0.6\\
G\,329.610+0.114$^{3}$		&	16 02 03.14	&	 --52 35 33.5	&	2008 Dec	&	20	&	--60.2	&	--61.4,--59.5	&	11	\\
G\,329.622+0.138			&	16 02 00.30	&	 --52 33 59.4	&	2008 Jun	&	0.5	&	--84.9	&	--85.0,--84.7	&	0.1	\\
						&				&				&	2008 Dec	&	$<$0.80	&		&	--200,140	&	\\
						&				&				&	2010 Mar	& 	$<$0.60	& & --150,190\\ 
G\,329.719+1.164			&	15 58 07.09	&	 --51 43 32.6	&	2008 Dec	&	1.9	&	--75.6	&	--75.9,--75.5	&	0.6	\\

\end{tabular}\label{tab:12MMB}
\end{table*}

\begin{figure*}
	\epsfig{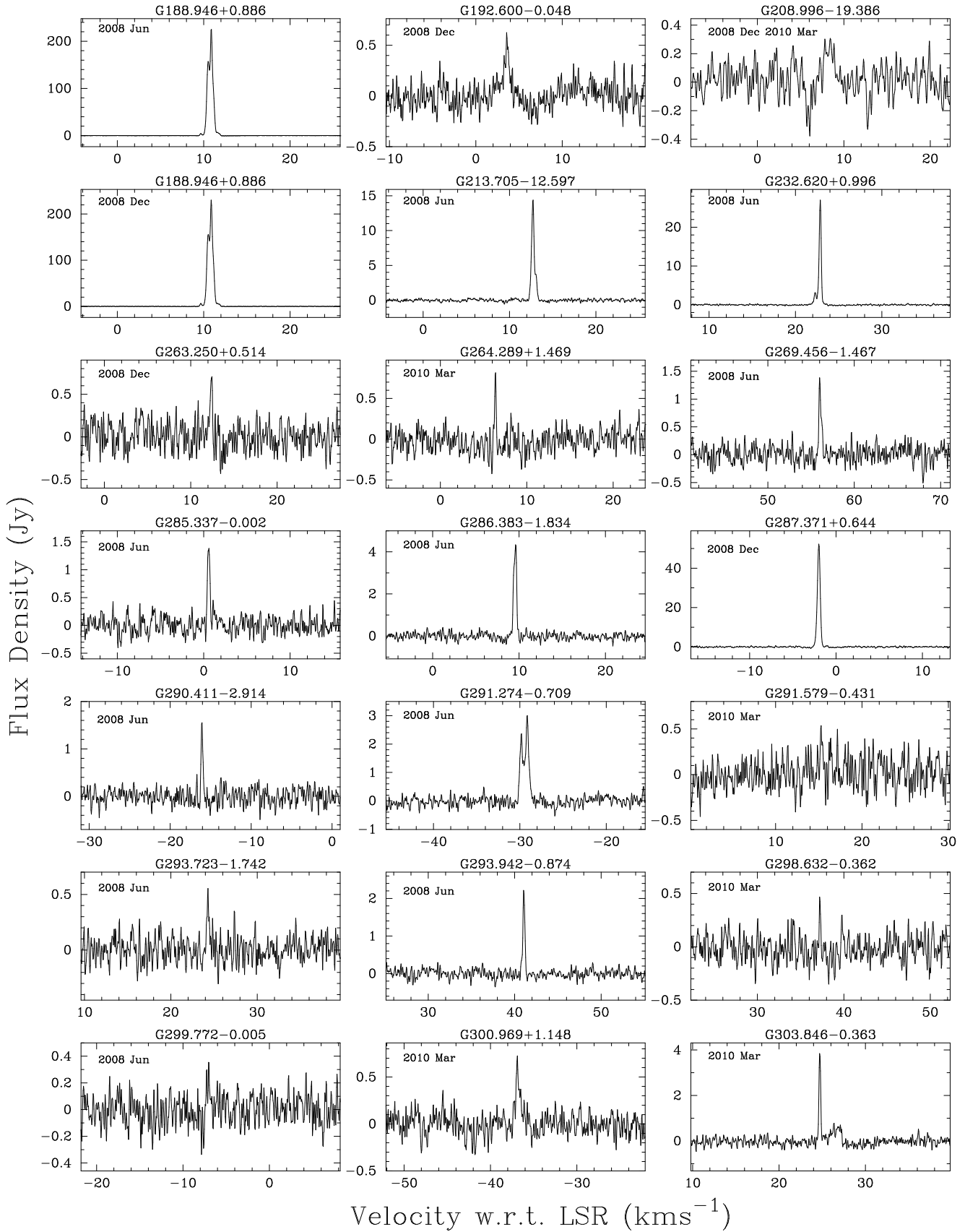}
\caption{Spectra of the 12.2-GHz methanol masers detected towards 6.7-GHz MMB sources.}
\label{fig:12MMB}
\end{figure*}

\begin{figure*}\addtocounter{figure}{-1}
	\epsfig{figure=12mmb_paper2_2.eps}
\caption{--{\emph {continuued}}}
\end{figure*}

\begin{figure*}\addtocounter{figure}{-1}
	\epsfig{figure=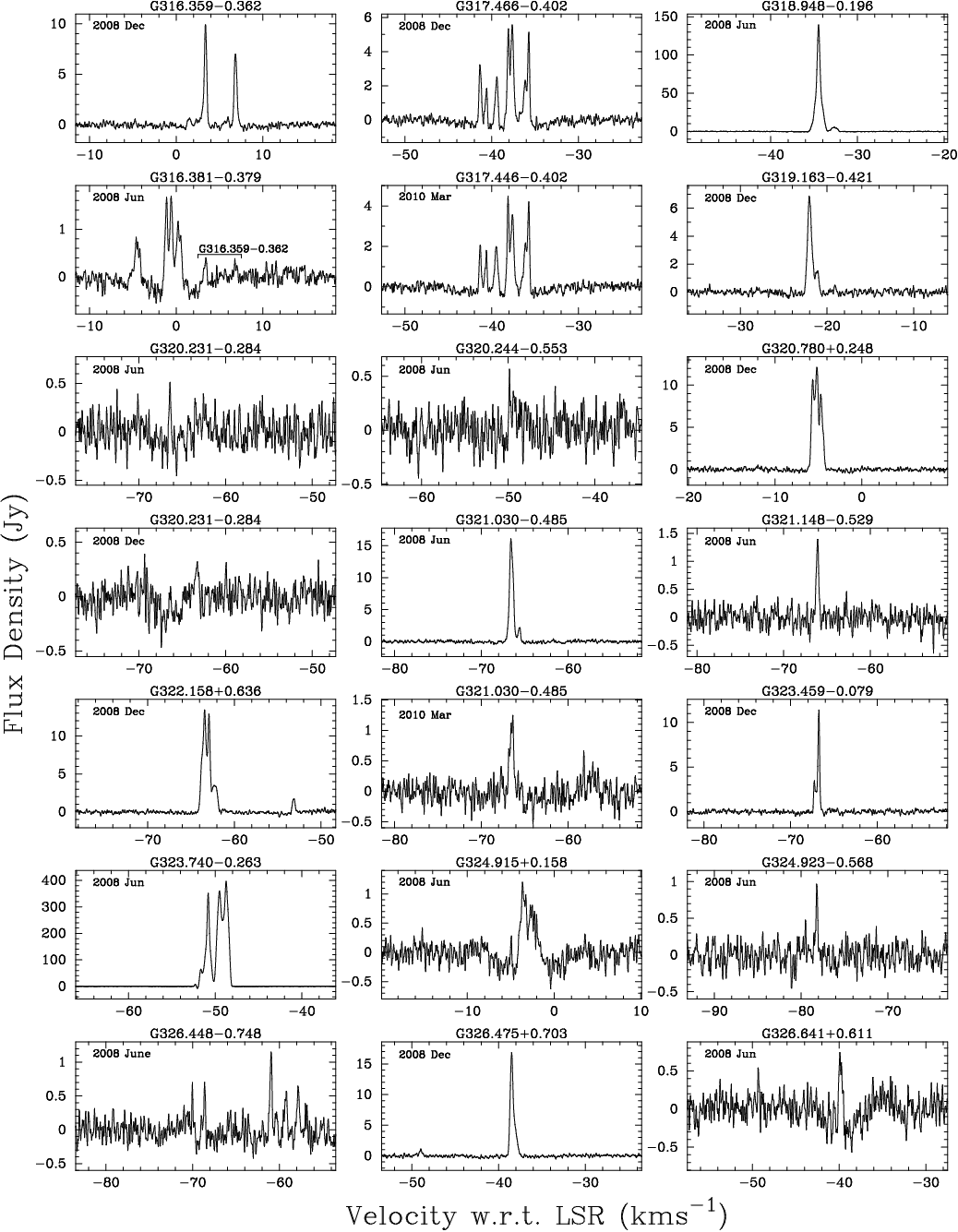}
\caption{--{\emph {continuued}}}
\end{figure*}

\begin{figure*}\addtocounter{figure}{-1}
	\epsfig{figure=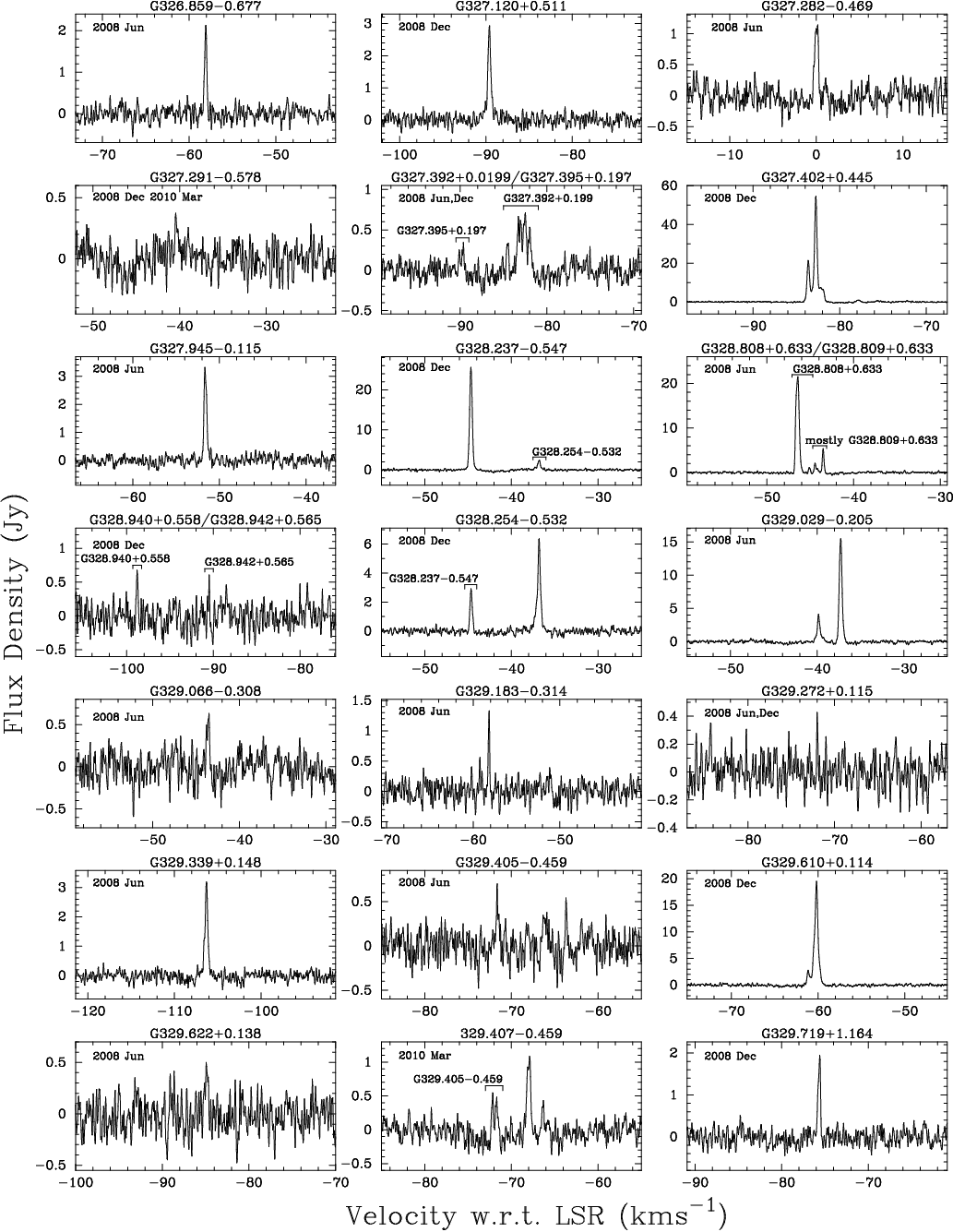}
\caption{--{\emph {continuued}}}
\end{figure*}

\section{Individual Sources}\label{sect:ind}

In this section we draw attention to complex sources, including information that cannot be adequately conveyed in Tables~\ref{tab:6MMB} and \ref{tab:12MMB}, such as those which are confused by emission from nearby sources, as well as those which are particularly variable. Historical remarks and associations with other maser species are often included. Instances where the flux density of the 12.2-GHz emission surpasses the 6.7-GHz emission are also noted.


\subparagraph{G\,188.946+0.886.} A well studied source (see Table~\ref{tab:12MMB} for references) that showed remarkably little variation between observations conducted during 2008 June and December and both spectra are presented in Fig.~\ref{fig:12MMB} to highlight this. This source has also changed little in peak flux density since the observations presented in \citet{Caswell95b}, when it was detected at 235~Jy, although the peak and secondary features have swapped roles since the earlier observations.

\subparagraph{G\,208.996--19.386.} 6.7-GHz emission was first detected towards this source within Orion A by \citet{Caswell95a} and later positioned by \citet{Voronkov05}. The ATCA observations of \citet{Voronkov05} partially resolved the 6.7-GHz emission, leading the authors to suggest that the emission might be thermal. MMB observations failed to detect any emission in either the survey, or follow-up MX observations \citep{Green12}.

We detect very weak emission at 12.2-GHz, covering a subset of the velocity range where 6.7-GHz emission is detected. Fig.~\ref{fig:12MMB} shows the average spectrum from the 2008 December and 2010 March, which has had an additional level of Hanning smoothing applied. The properties of the detected 12.2-GHz emission are consistent with it being of a thermal nature.

%

\subparagraph{G\,212.06--0.74.} No Parkes observations were conducted towards this source, instead we present a modest upper limit from observations using the Hobart 26-m telescope. The position of this 6.7-GHz methanol maser was not derived in the follow-up MMB observations and instead the Effelsberg position of \citet{Xu08} is presented in \citet{Green12} and used as our 12.2-GHz target position.

\subparagraph{G\,196.454--1.677 and G\,284.352--0.419.} 12.2-GHz emission was detected towards these sources by \citet{Caswell95b} at a peak flux densities of 1.35~Jy and 0.6~Jy, respectively. We detected no emission during either of two observations conducted towards each of these sources. This is consistent with variability: we find that 14 sources whose peak flux density was no more than 1.5~Jy when detected, dropped below our detection limit on at least one of the observing epochs (see Section~\ref{sect:var} for details).

\subparagraph{G\,292.074--1.131.} The epoch-averaged spectrum towards this source shows possible marginal emission ($\sim$0.3~Jy) at a velocity of --19.0 \kmsns, consistent with the 0.85~Jy peak of the 6.7-GHz methanol maser at --19.1 \kms \citep{Green12}. Due to the similar strength of nearby noise spikes, we do not claim this source as a detection, and further observations are required to determine the authenticity of the emission.

\subparagraph{G\,294.977--1.734.} 6.7-GHz methanol maser emission was detected towards this source in the MMB survey observations, but not in the follow-up MX observation to a 5-$\sigma$ sensitivity of 0.1~Jy. No 12.2-GHz methanol maser emission was detected in either of our 2008 June or December observations. Since this source is clearly variable at 6.7-GHz, further observations would be required at 12.2-GHz to rule out any variable emission at this transition.

\subparagraph{G\,299.772--0.005.} While this source is only detected at a 3.6-$\sigma$ level during 2008 June, it is clearly visible at the velocity of the 6.7~GHz source peak. No emission was detected in either 2008 December or 2010 March.

\subparagraph{G\,305.202+0.208 and G\,305.208+0.206.} These nearby sources were observed during both 2008 June and December. G\,305.208+0.206 showed little variation between the two epochs, while G\,305.202+0.208 halved in flux density between the observations in June and those in December (from 3.5 to 1.8~Jy). G\,305.208+0.206 is associated with both OH and water masers \citep{C98,Breen10b}.

\subparagraph{G\,305.366+0.184.} This 12.2-GHz source is associated with a 6.7~GHz secondary feature and shows stronger emission at 12.2 than 6.7~GHz (see Fig.~\ref{fig:12stronger}). The 6.7~GHz observations were carried out on 2008 March 13, just over three months prior to the 12.2-GHz observations. While it is possible that variability between the epochs can account for this unusual circumstance, this source is worthy of concurrent follow-up observation at both frequencies.


\subparagraph{G\,308.918+0.123.} 12.2-GHz observations during 2008 December showed stronger emission than was present at 6.7-GHz in 2009 March 12, almost three months later. Like G\,305.366+0.184, this source needs further observations at both frequencies to determine if this is due to variability. This source exhibits both OH and water maser emission \citep{C98,Breen10b}.

\subparagraph{G\,317.466--0.402.} The 12.2-GHz emission (observed 2008 December) is stronger than its 6.7-GHz counterpart (observed 2008 Aug 19) for two features between --36.5 and --35.6 \kmsns. Fig.~\ref{fig:12MMB} shows spectra from both 2008 December and 2010 March, showing typical moderate levels of variability between the two epochs. 

\subparagraph{G\,319.163--0.421.} The peak of the 12.2-GHz emission is associated with a secondary feature in the 6.7-GHz spectrum and is stronger. The 6.7-GHz observations were carried out during 2008 August 20, four months prior to the 12.2-GHz follow-up observations.

\subparagraph{G\,320.231--0.284.} Spectra of the 12.2-GHz emission from both 2008 June and December are shown in Fig.~\ref{fig:12MMB}, allowing the variability between the two epochs to be seen. Weak emission was detected at both epochs from different features, showing variability of at least a factor of two. Both OH and water masers are also coincident with this methanol maser site \citep{C98,Breen10b}.

\subparagraph{G\,321.030--0.485 and G\,321.033--0.483.} This close pair of 6.7-GHz methanol masers have overlapping velocity ranges and many confused features, making it impossible to accurately determine integrated flux densities. The velocity ranges of the two sources have been taken from the \citet{Green12} ATCA observations. The 12.2-GHz methanol maser emission we detect is at the velocity of the peak emission attributed to G\,321.030--0.485 and thus we expect the 12.2-GHz emission to be located here. It is possible that some of the 12.2-GHz methanol maser emission is associated with G\,321.033--0.483, but high resolution observations would be required to make a confident conclusion. The 12.2-GHz methanol maser emission displays the most extreme variation of all the observed sources, reducing from 16~Jy during 2008 June, to 1.2~Jy in 2010 March. Spectra from both epochs are shown in Fig.~\ref{fig:12MMB}.

\subparagraph{G\,321.704+1.168.} 6.7-GHz methanol maser emission was detected here in the MMB survey observations (and in the earlier observations of \citet{Caswell95a}), but not in the follow-up MX observation to a 5-$\sigma$ sensitivity of 0.1~Jy. No 12.2-GHz methanol maser emission was detected in either of our 2008 June or December observations. Since this source is clearly variable at 6.7-GHz, further observations would be required at 12.2-GHz to rule out any variable emission in this transition.

\subparagraph{G\,326.641+0.661.} The 12.2-GHz emission detected in 2008 June consisted of a single spectral feature at a velocity of --39.9 \kms\, with a peak flux density of 0.7~Jy. While this emission is just 5-$\sigma$, the main peak extends beyond 0.3~Jy in at least 8 adjacent spectral channels. Observations of the target 6.7~GHz methanol maser during 2008 March 18 showed no emission at this velocity. A spectrum showing both transitions is presented in Fig.~\ref{fig:12weird}. 

\subparagraph{G\,328.808+0.633 and G\,328.809+0.633.} This close pair of sources are separated by less than 2 arcsec, making the emission very difficult to separate. While emission at 12.2-GHz between --46.9 and --44.6 \kms can be uniquely attributed to G\,328.808+0.633 (on the basis of velocity ranges seen in high resolution 6.7-GHz observations), some component of the emission between --44.6 and --43.4 \kms is likely to be present at both positions with perhaps the strongest contribution from G\,328.809+0.633. This region exhibits a wealth of masers, including OH and water \citep{C98,Breen10b}, but also many additional methanol and excited OH transitions, including: 19.9, 85.5 and 107~GHz methanol \citep{Ellingsen04,Ellingsen03,V+99} as well as 1720, 4765, 6030, 6035 and 13441~MHz excited OH emission \citep{DE02,C03,Cas04b,Cas04c}.

\section{Discussion}

This second portion of our 12.2-GHz methanol maser MMB follow-up catalogue forms part of the largest sensitive search of these methanol masers to date. Furthermore, given that the observations target a complete sample of 6.7-GHz methanol masers from the MMB survey, this project presents the most complete catalogue of 12.2-GHz methanol masers to date. The current catalogue, encompassing a longitude coverage of 186$^{\circ}$ to 330$^{\circ}$ complements the already published longitude range of 330$^{\circ}$ (through 360$^{\circ}$) to 10$^{\circ}$ \citep{BreenMMB12}. The global properties of 12.2-GHz methanol masers detected towards the full sample of MMB sources that lie south of declination --20$^{\circ}$ (580 sources) has also been given \citep{Breen12stats}.

The analysis presented in \citet{Breen12stats,BreenMMB12} confirmed that the detected 12.2-GHz masers were chiefly observed towards the stronger 6.7-GHz methanol maser targets. The detections and source properties are consistent with the notion presented previously that 12.2-GHz methanol masers are generally associated with the second half of the 6.7-GHz methanol maser lifetime, and an evolutionary scenario whereby both the flux density and velocity range of the maser emission increase as the source evolves. Consideration of the luminosity ratios of individual features associated with 20 methanol maser sites showing both transitions revealed further support for the idea that the luminosities of the sources change as a function of evolution \citep{BreenMMB12}.

In \citet{BreenMMB12} we investigated the completeness of this 12.2-GHz search by stacking 12.2-GHz non-detections after first aligning the spectra at the velocity of the peak 6.7-GHz methanol maser emission, which works well since 80 per cent of 12.2-GHz methanol masers share their peak velocity with their 6.7-GHz counterpart \citep{Breen12stats}. From this analysis, \citet{BreenMMB12} concluded that the 12.2-GHz non-detections have no emission above 0.09~Jy, on average, confirming that there are few 12.2-GHz methanol masers that fall just below our detection limit.

In the final paper of this series we will extend the 12.2-GHz methanol maser follow-up catalogue to longitudes greater than 10$^{\circ}$. Here we focus on the specific data and unusual sources of special interest that fall in the longitude range of 186$^{\circ}$ to 330$^{\circ}$.

\subsection{Basic Statistics}

In the longitude range 186$^{\circ}$ to 330$^{\circ}$ we detect 12.2-GHz methanol masers towards 40 per cent of the target 6.7-GHz methanol masers. In comparison, we detected emission at 12.2-GHz towards 46 per cent of the 6.7-GHz methanol methanol masers in the longitude range 330$^{\circ}$ to 10$^{\circ}$ \citep{BreenMMB12}. The slightly higher 12.2-GHz detection rate in the 330$^{\circ}$ to 10$^{\circ}$ can be attributed to its inclusion of the 335$^{\circ}$ to 340$^{\circ}$ longitude range, noted by \citet{Breen12stats} as having a statistically significantly higher 12.2-GHz detection rate than other parts of the Galactic plane.

As was the case in \citet{Breen12stats,BreenMMB12}, the 12.2-GHz detection rate achieved in the presented longitude range is lower than previous large, sensitive surveys \citep[e.g.][detection rates of 55 and 60 per cent, respectively]{Caswell95b,Breen10a} and likewise less sensitive surveys (once the different detection limits have been accounted for) such as \citet{Gay,Blas04}. We attribute this detection rate discrepancy to the sample biases that affected previous searches, which have preferentially focused on more evolved star formation regions: targeting either 6.7 GHz methanol masers that had been detected towards OH maser targets, or 6.7 GHz methanol masers detected towards IRAS selected regions. Our systematic search towards a complete sample of 6.7-GHz methanol masers encompasses all evolutionary stages traced by 6.7-GHz methanol masers, importantly including the youngest sources. \citet{Breen10a} find that 12.2-GHz methanol masers are generally present towards the second half of the 6.7-GHz maser lifetime, accounting for the higher detection rates achieved in searches  biased towards more evolved objects.

In the 186$^{\circ}$ to 330$^{\circ}$ longitude range we detect 12.2-GHz methanol masers that range in flux density from 0.3~Jy (G\,327.395+0.197) to 396~Jy (G\,323.740-0.263), and exhibit emission over 0.1 \kms to 13.1 \kmsns. Many of the strongest sources detected are known; the average flux density of the known 12.2-GHz sources is 31~Jy, while the average of the newly detected sources is 2.6~Jy (the average of the total sample is 18~Jy). The strongest newly detected source is G\,326.475+0.703 at 17~Jy.

\begin{figure}
	\epsfig{figure=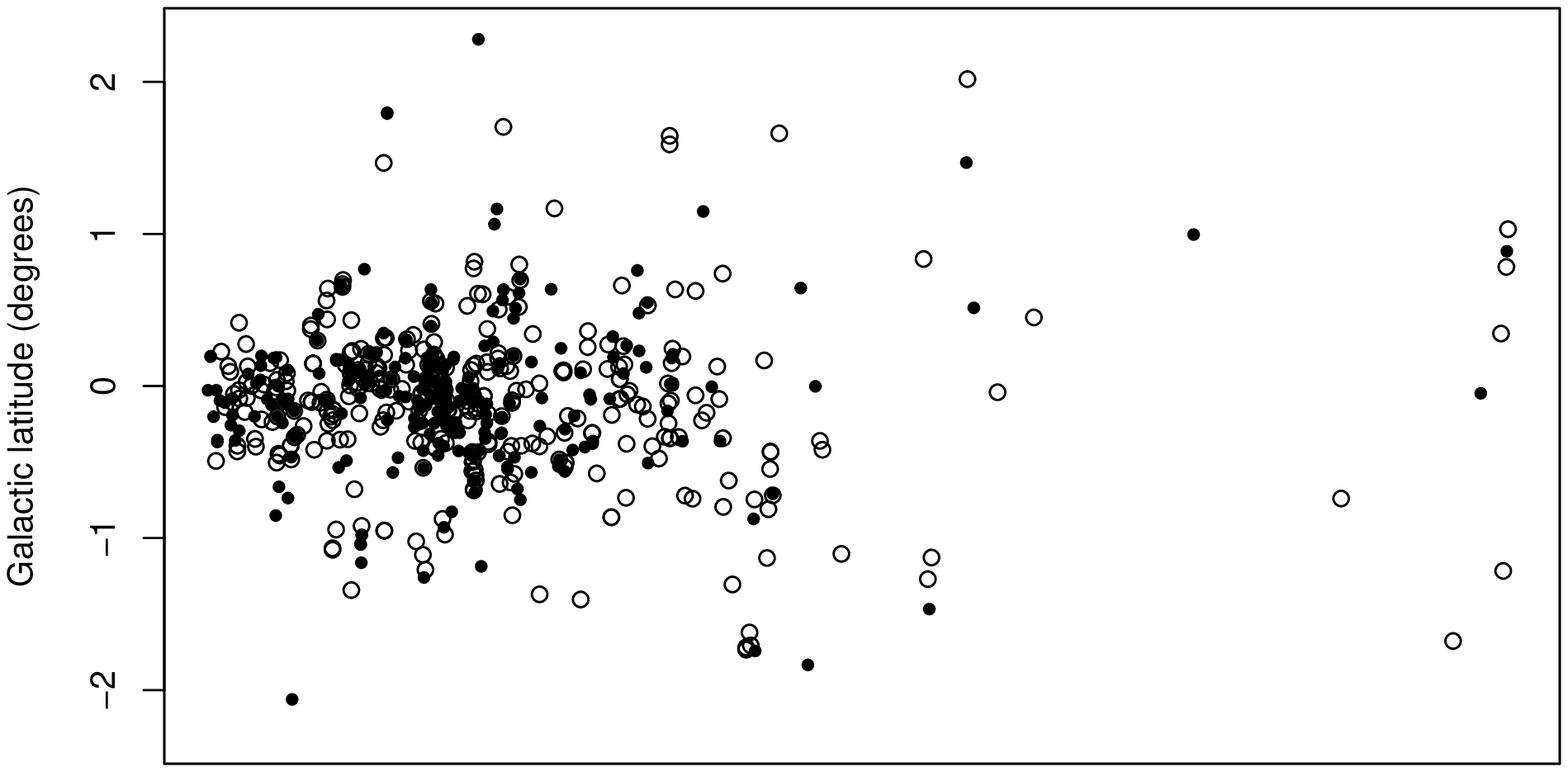,angle=270,width=9cm}\vspace{-2.1cm}
	\epsfig{figure=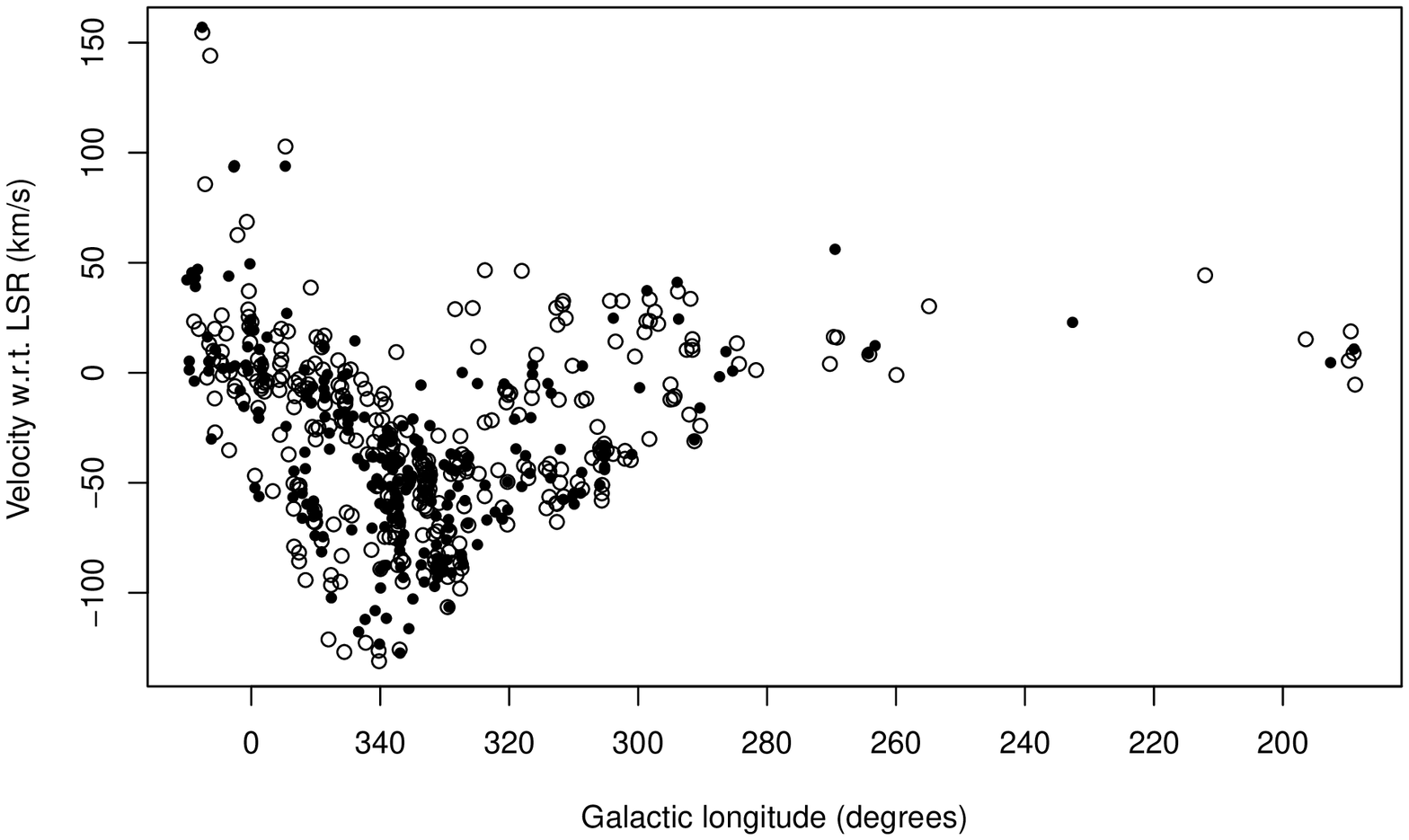,angle=270,width=9cm}
\caption{Distribution of the methanol masers in the Galactic longitude range 186$^{\circ}$ (through 360$^{\circ}$) to 10$^{\circ}$ (top) and distribution of the peak 6.7-GHz methanol maser velocity versus Galactic longitude (bottom). Dots mark the locations of 6.7-GHz methanol masers with 12.2-GHz counterparts, while open circles show those 6.7-GHz sources devoid of 12.2-GHz emission.}
\label{fig:dist}
\end{figure}

The distribution of the 6.7-GHz methanol masers throughout the Galactic longitude ranges 186$^{\circ}$ (through the Galactic centre) to 10$^{\circ}$, wholly incorporating the longitude ranges given in the current paper, together with that in \citet{BreenMMB12}, is shown in Fig.~\ref{fig:dist}. The second panel in this figure shows the distribution of methanol masers as a function of line-of-sight velocities. In both panels, 6.7-GHz methanol masers with accompanying 12.2-GHz emission have been distinguished from those sources exhibiting no 12.2-GHz methanol maser emission.

%
%
%

The longitude-velocity distribution shows 12.2-GHz masers associated with the innermost part of the Galaxy, the Galactic centre region described by \citet{CasMMB10}. Four sources with high positive velocities ($\geq$ 100 \kmsns) between $\pm$10 degrees longitude are likely to be associated with the Galactic bar(s), which represents approximately half the 6.7-GHz methanol population. There are relatively few 12.2-GHz masers detected towards 6.7-GHz sources associated with the 3-kpc arms (~25 per cent, compared with 40 per cent for full longitude range presented here), perhaps indicating that the high-mass star formation regions in this feature of the Galaxy are generally younger than other regions in the Galaxy. The high density of sources between longitudes 330$^{\circ}$ and 340$^{\circ}$ remains prominent and would be coincident with the tangent of a continuous 3 kpc arm structure and the origin of the Perseus spiral arm \citep{Green11} and there is also a cluster of sources coincident with the Carina-Sagittarius origin and bar interaction with the 3-kpc arms. \citet{Breen12stats} found that the 335$^{\circ}$ to 340$^{\circ}$ longitude range also has a significantly higher 12.2-GHz detection rate (after accounting for the high density of 6.7-GHz sources in this longitude range). There are relatively fewer detections towards the 6.7-GHz methanol masers associated with the outer Galaxy portions of the Carina-Sagittarius and Crux-Scutum spiral arms is at least partially attributable to sensitivity limitations.



\subsection{Variability}\label{sect:var}

The sparseness of our observing epochs prevents a thorough investigation of intrinsic 12.2-GHz maser variability. However, we are able to recognise several interesting properties from our observations which observed individual sources up to three times over a period of up to $\sim$20 months. Multiple repeat observations were preferentially made towards sources where either no, or marginal, emission was detected during the previous epoch.

Only 28 of our 83 12.2-GHz detections were observed with a peak flux density of more than 2~Jy at any of the observation epochs. Peak flux density variations of these sources, which are less susceptible to large relative variations due to noise in the respective spectra, show that in general, the stronger sources are less variable. 13 of this sample of 28 sources show total peak flux density variations of less than 20 per cent over the course of their observation and have an average maximum peak flux density of 56~Jy. In comparison, the 15 sources which exhibit variations in excess of 20 per cent have an average peak flux density of 8.4~Jy, and an average percentage variation of 42 per cent. The most extreme variation is shown by G\,321.030--0.485 which reduced from 16 to 1.2~Jy over a 20 month period (both spectra are shown in Fig.~\ref{fig:12MMB}).

Further evidence that weaker 12.2-GHz masers may exhibit greater variability lies in their detectability over the multiple observation epochs. Of the 83 12.2-GHz methanol masers that we detect, four were identified only in the multiple epoch average spectrum, and of the remaining 79, 14 (or 18~per cent) could not be identified during at least one observation epoch. All of these 14 sources belong to the group of sources with flux densities of no more that 1.5~Jy at any observing epoch and were observed on at least two separate observing sessions; a group with 27 members in total. All 12.2-GHz sources that were observed with a peak flux density greater than 1.5~Jy on any occasion remained detectable during each of the observations. 

Our variability findings are consistent with earlier remarks made by \citet{Caswell95b} that 12.2-GHz methanol masers commonly show moderate levels of variations, but rarely the extreme levels that have been observed in water masers. \citet{Felli07} conducted monitoring observations towards 43 water maser sites over a period of 20 years, showing the enormity of the variability that can occur over various timescales. In a much shorter timescale study, \citet{Breen10b} conducted sensitive water maser observations at two epochs separated by $\sim$10 months and found that a number of sources with flux densities greater than 2~Jy when detected fell below their detection limit (5-$\sigma$ of 0.2~Jy) at the other epoch, the most extreme of which was 80~Jy when detected.

12.2-GHz methanol masers have been detected previously towards G\,196.454--1.677 and G\,284.352--0.419 \citep{Caswell95b} with flux densities of 1.35 and 0.7~Jy, respectively, but, in our observations we failed to detect any 12.2-GHz methanol maser emission towards either source to 5-$\sigma$ detection limits of 0.75 and 0.55~Jy during the 2008 June and December observations respectively. The non-detection of both these sources is entirely consistent with the level of variability we see in the full sample of these weak 12.2-GHz methanol masers.


%

\subsection{6.7-GHz methanol masers with stronger 12.2-GHz features}

\begin{figure*}
	\epsfig{figure=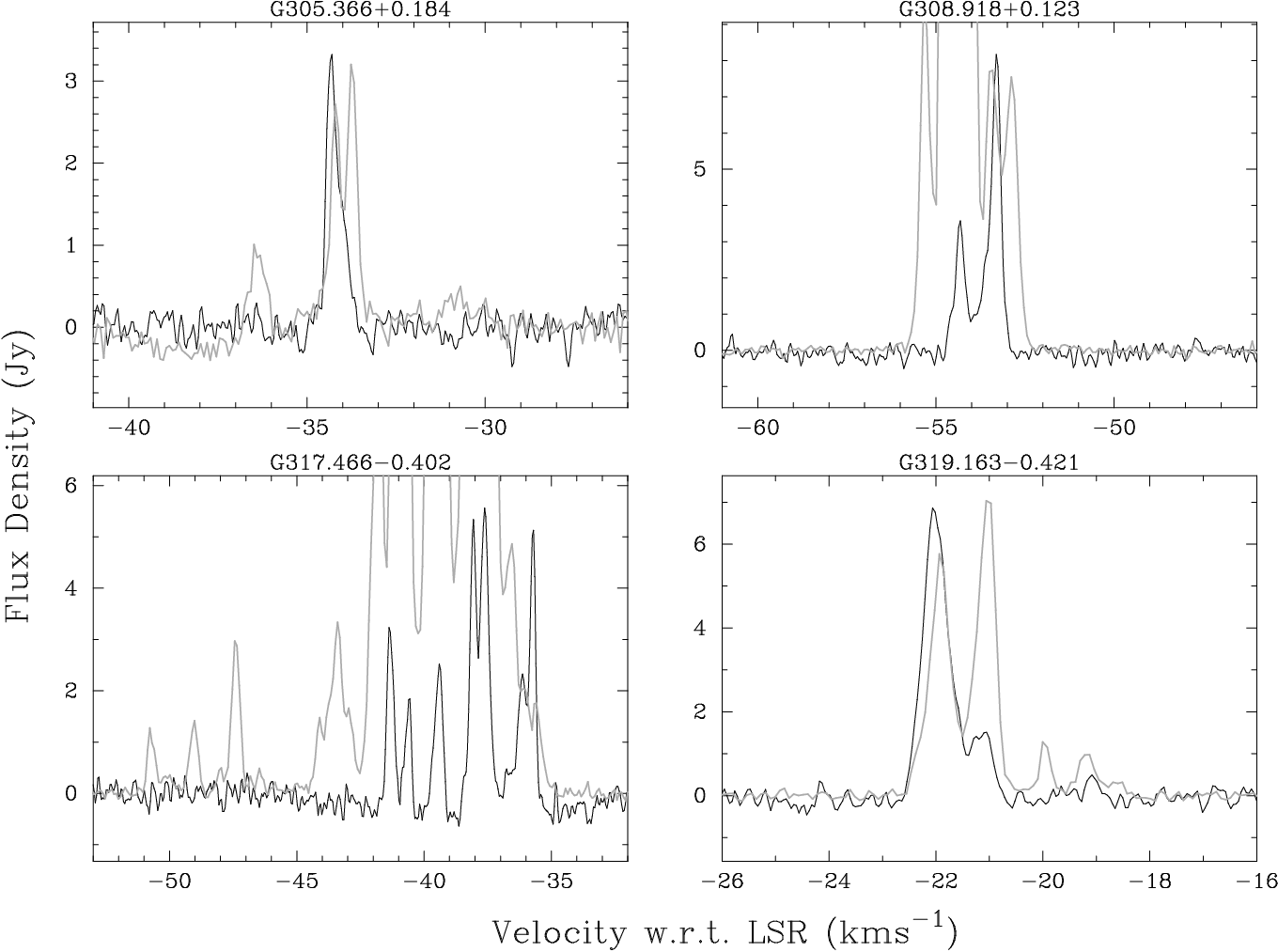,angle=270,width=16cm}
\caption{Comparison between the 12.2-GHz (black) and 6.7-GHz emission (grey) for sources showing stronger emission at 12.2-GHz for at least one spectral feature.}
\label{fig:12stronger}
\end{figure*}

The flux density of 12.2-GHz methanol maser emission rarely surpasses the flux density of the associated 6.7-GHz methanol maser emission \citep[e.g.][]{Caswell95b,BreenMMB12}. The maser modelling of \citet{Cragg05} shows that while having emission from the two transitions of equal brightness temperature is very rare (i.e. 12.2-GHz emission 3.3 times larger in flux density than the 6.7-GHz counterpart), similar flux densities are more likely, although observationally uncommon. In the 330$^{\circ}$ to 10$^{\circ}$ longitude portion of the catalogue we found 11 12.2-GHz sources with at least one feature that surpassed their 6.7-GHz emission in flux density. Here we find a similar portion of sources with stronger 12.2-GHz features (4 of 83) and their 6.7- and 12.2-GHz spectra are presented in Fig.~\ref{fig:12stronger}.

Fig.~\ref{fig:12stronger} shows one source of particular interest: G\,317.446--0.402. This source exhibits a 12.2-GHz feature of almost three times the flux density of the coincident 6.7-GHz emission (at about --36 \kmsns). The emission at the two transitions therefore have approximately equal brightness temperatures, close to the most extreme predicted ratios reported by the maser models \citep{Cragg05}.

\begin{figure}
\begin{center}
	\epsfig{figure=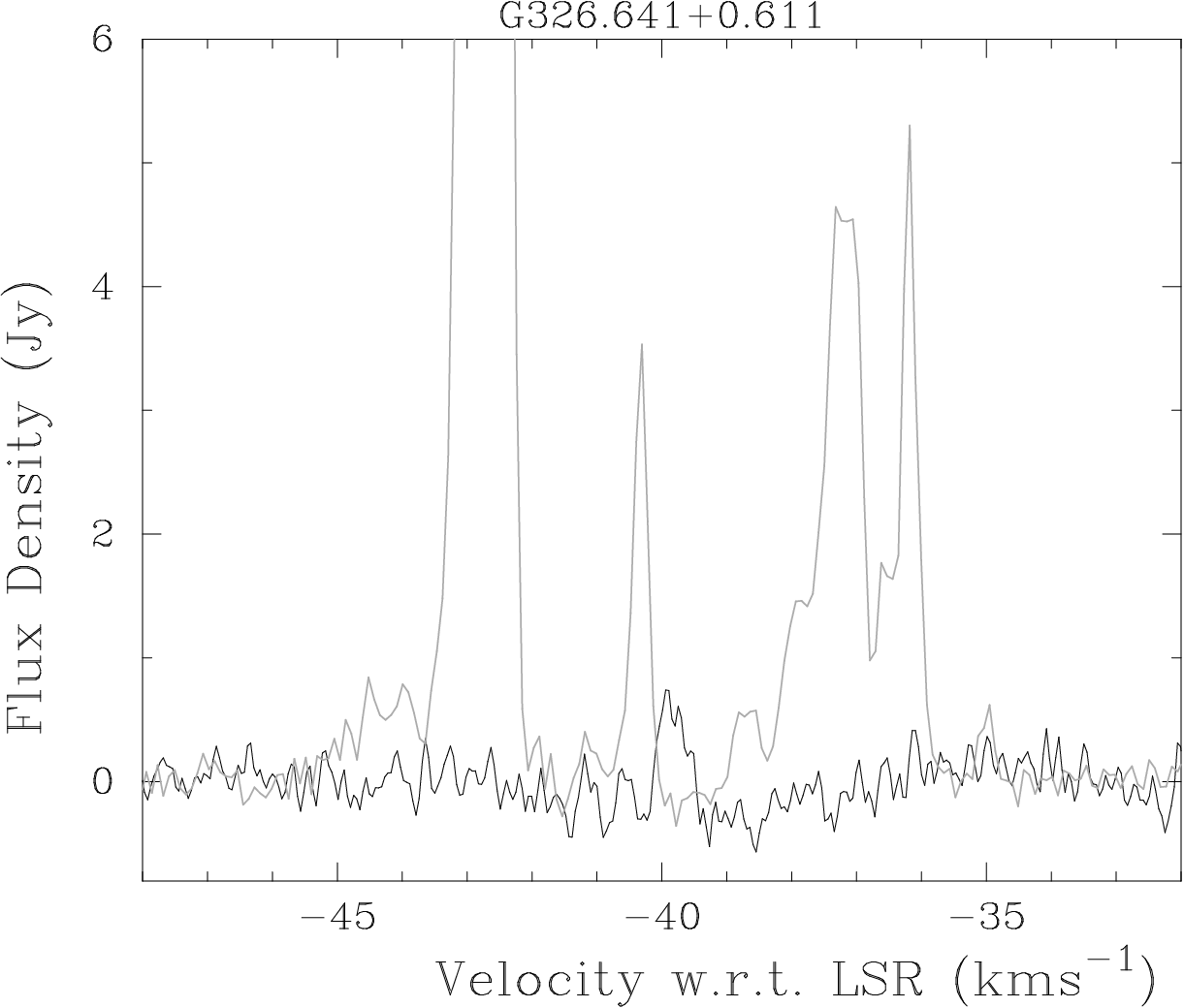,height=8cm,angle=270}
\caption{Comparison between the 12.2-GHz emission (black) and 6.7-GHz emission (grey) for G\,326.641+0.611. The deviation in the 12.2-GHz baseline is probably due to real absorption. Note that the uncertainties in the adopted rest frequencies correspond to velocities of 0.1~\kms and 0.03~\kms for the 12.2-GHz and 6.7-GHz transitions, respectively \citep{Muller04}.}
\label{fig:12weird}
\end{center}
\end{figure}

We detect a further source, G\,326.641+0.611, which apparently exhibits 12.2-GHz emission at a velocity with no detected 6.7-GHz emission (Fig.~\ref{fig:12weird}). The characteristics of this sources are even more unusual and require further, near contemporaneous observations at both frequencies to investigate its nature. 

In the current longitude range, 2 per cent of 6.7-GHz masers exhibit at least one 12.2-GHz maser feature with a larger flux density, consistent with the 3 per cent of sources in the previous portion of the catalogue (and 2.5 per cent of the combined catalogue). \citet{BreenMMB12} concluded that the sources showing stronger 12.2-GHz emission did not share any obvious common properties, such as similar luminosities or association with other maser species, and are therefore not associated with a distinct phase of source evolution.  

\subsection{Targeting rare methanol maser transitions}

In a survey for 107.0-GHz and 156.6-GHz methanol masers, \citet{Caswell00} found that there was a tendency for 107.0-GHz methanol masers to be associated with the 6.7- and 12.2-GHz methanol masers with the largest flux densities. Recent comparisons of the luminosity of 6.7- and 12.2-GHz methanol masers by \citet{Ellingsen11} showed that this is a luminosity relationship, with the occurrence of 107.0-GHz methanol masers intimately tied to the luminosity of the associated 6.7-GHz methanol maser emission.

\begin{figure}\vspace{-1.5cm}
\begin{center}
	\epsfig{figure=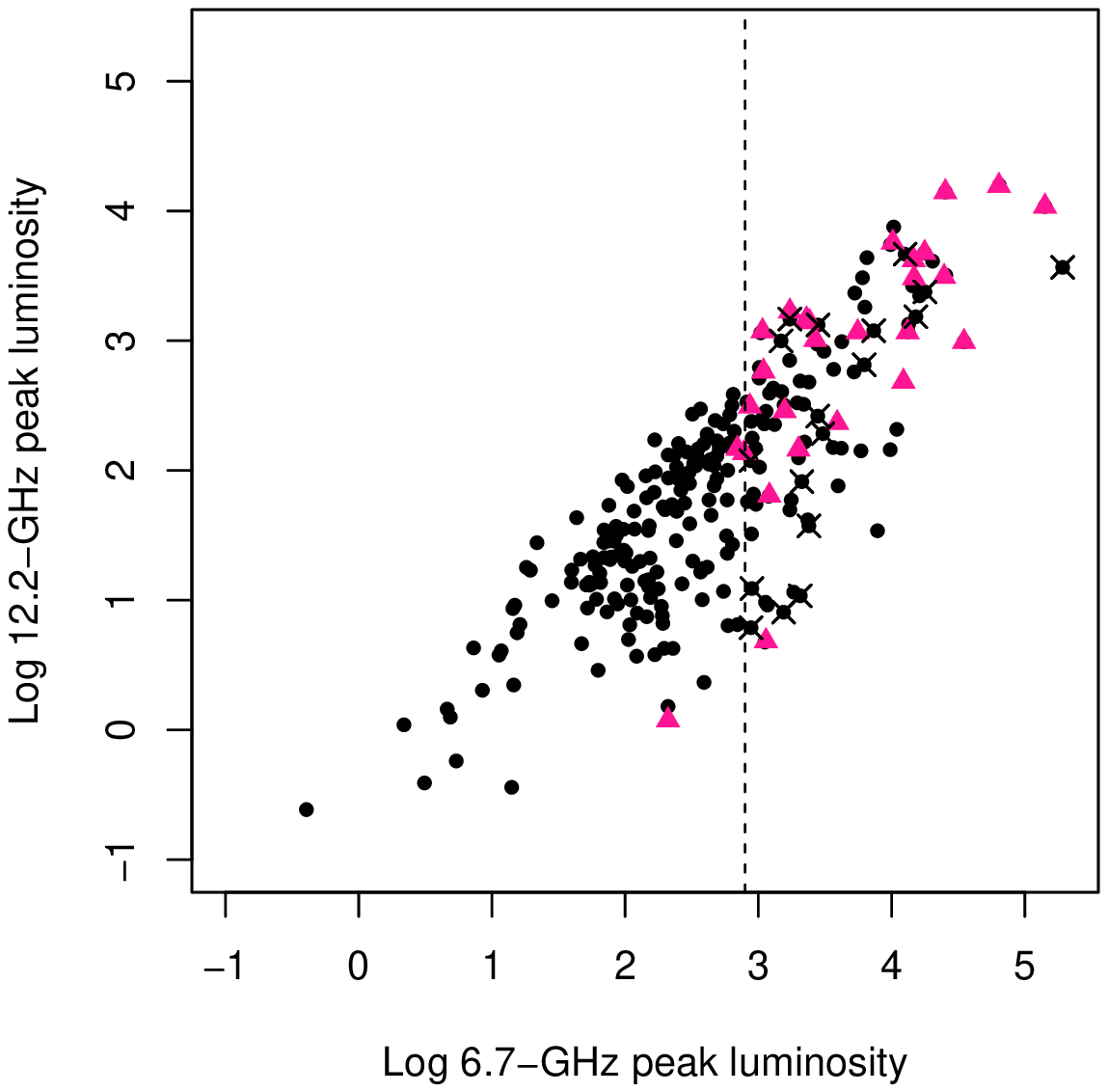,height=9cm,angle=270}
\caption{Peak luminosity of the 12.2-GHz maser vs the peak luminosity of the 6.7-GHz maser peak (units of Jy kpc$^2$) for 241 of the 267 12.2 GHz methanol masers detected in the longitude range 186$^{\circ}$ (through the Galactic centre) to 10$^{\circ}$  \citep[][plus the current paper]{BreenMMB12} (black dots). The pink triangles represent sources with an associated 107.0-GHz methanol maser \citep{V+99,Caswell00} and the vertical dashed line at 2.9 shows the line that \citet{Ellingsen11} suggests approximates the cutoff for the presence of 107 GHz masers. Therefore black dots to the right of this line are prime candidates for follow-up observations at 107.0-GHz. Sources that have been searched with no emission detected have a black cross through them. }
\label{fig:107}
\end{center}
\end{figure}

Fig.~\ref{fig:107} has been adapted from fig.\ 4 of \citet{Ellingsen11} and shows the luminosity of the peak 12.2-GHz emission plotted against the luminosity of the peak 6.7-GHz methanol maser emission. The peak flux densities used in this plot have been taken from the observations reported in the current paper along with those in \citet{BreenMMB12} and MMB data. Luminosities have been calculated assuming isotropic emission and for distances we have used parallax measurements or H{\sc i} self-absorption ($\sim$70 per cent of sources) where available and assumed the near kinematic distance (calculated using the prescription of \citet{Reid09}) for the remaining sources. This figure shows a clear tendency for the 107.0-GHz methanol masers (marked with pink triangles) to be associated with the most luminous 6.7-GHz methanol masers. \citet{Ellingsen11} suggested a 6.7-GHz luminosity cutoff where 107.0-GHz methanol masers were likely to be found and this is marked by a vertical dashed line in Fig.~\ref{fig:107}.

\citet{Ellingsen11} suggested that methanol masers that fell to the right of their 6.7-GHz luminosity cutoff would make excellent targets for future 107.0-GHz observations. We have marked those 6.7-GHz methanol masers that have been searched, but failed to reveal any 107.0-GHz emission \citep{V+99,Caswell00} on Fig.~\ref{fig:107} with black crosses. The sources corresponding to the remaining 45 black dots in this region are listed in Table~\ref{tab:107}. If we assume that the detection rate for 107.0-GHz methanol masers associated with 6.7-GHz masers with peak luminosity $>$ 790 Jy kpc$^2$ is 42 per cent \citep[as observed by ][]{Ellingsen11}, then we would expect to detected approximately twenty 107.0-GHz methanol masers by targeting these candidates.  The 107.0-GHz methanol masers typically have peak flux densities around an order of magnitude less than the associated 6.7 GHz masers.  Looking at Table~1 in this paper and \citet{BreenMMB12}, the majority of sources listed in Table~\ref{tab:107} have 6.7-GHz peak flux densities of 10s of Jy, so we would predict the 107.0~GHz peak flux densities in these sources are likely to be $<$ 10~Jy.  Sites of 107.0-GHz methanol maser emission have traditionally made excellent targets for other more rare methanol maser transitions such as those at 19.9-, 23.1, 37.7- and 38.3-GHz \citep{Cragg03,Ellingsen04,Ellingsen11}.

Methanol masers at 6.7- and 12.2-GHz have proven to be excellent evolutionary probes \citep[e.g.][]{Breen12stats,Breen10a} following a scenario whereby 12.2-GHz methanol masers are present towards the later half of the 6.7-GHz methanol maser lifetime. \citet{Breen12stats} showed that the properties of these masers could offer even finer evolutionary detail, expecting that these masers increase in luminosity as they evolve, albeit at a slower rate at 12.2-GHz. \citet{Ellingsen11} used this relationship to show that the presence of some rarer methanol masers, including the 107.0-GHz transition, are present towards the end of the 6.7- and 12.2-GHz methanol maser phase.  Observations of multiple methanol maser transitions in the same source can place strong constraints on theoretical models of the maser emission, which in turn can be used to investigate the physical conditions in these regions at extremely high resolution.  This approach has been demonstrated in a small number of sources \citep[e.g.][]{Cragg01,Sutton01}, but the new capabilities of the Australia Telescope Compact Array (ATCA), and ALMA, will make it possible to undertake such studies much more easily, with higher spatial resolution, greater sensitivity and (near-)simultaneous coverage of more transitions.  Hence it is timely to identify a larger (and more representative) sample of potential targets for such multi-transition methanol maser studies.



\begin{table}
\caption{Sources with peak 6.7-GHz luminosities placing them to the right of the vertical dashed line in Fig.~\ref{fig:107}, making them excellent candidates for follow-up observations at 107.0-GHz.} 
  \begin{tabular}{lllllccclclllclllcll} \hline
\multicolumn{1}{c}{\bf Methanol maser}  &\multicolumn{1}{c}{\bf Methanol maser} & \multicolumn{1}{c}{\bf Methanol maser} \\
    \multicolumn{1}{c}{\bf ($l,b$)}   &\multicolumn{1}{c}{\bf ($l,b$)} &   \multicolumn{1}{c}{\bf ($l,b$)} \\
    \multicolumn{1}{c}{\bf (degrees)}  &  \multicolumn{1}{c}{\bf (degrees)}  &  \multicolumn{1}{c}{\bf (degrees)} \\  \hline \hline

G\,303.846--0.363	& 	G\,337.201+0.114	& 	G\,350.299+0.122\\
G\,313.469+0.190	&	G\,337.844--0.375	&	G\,350.344+0.116	\\
G\,317.466--0.402	& 	G\,338.561+0.218	&	G\,351.688+0.171	\\
G\,329.066--0.308	&	G\,338.926+0.634	&	G\,352.083+0.167	\\
G\,329.339+0.148	& 	G\,339.476+0.185	&	G\,354.496+0.083	\\
G\,331.425+0.264	&	G\,339.762+0.054	&	G\,357.967--0.163	\\
G\,331.556--0.121	&	G\,339.949--0.539	&	G\,358.371--0.468	\\	
G\,332.813--0.701	&	G\,339.986--0.425	&	G\,358.721--0.126	\\
G\,333.562--0.025	&	G\,341.276+0.062	&	G\,359.436--0.104	\\
G\,333.646+0.058	&	G\,342.484+0.183	&	G\,0.092--0.663		\\
G\,335.556--0.307	&	G\,343.354--0.067	&	G\,0.315--0.201		\\
G\,336.864+0.005	&	G\,343.929+0.125	&	G\,6.189--0.358		\\
G\,336.358--0.137	&	G\,346.480+0.221	&	G\,6.795--0.257		\\
G\,337.052--0.226	&	G\,347.863+0.019	&	G\,8.832--0.028		\\	
G\,337.097--0.929	&	G\,348.884+0.096	&	G\,9.986--0.028 	\\ \hline

\end{tabular}\label{tab:107}

\end{table}

\section{Summary}

We present the second instalment of our 12.2-GHz methanol maser observations targeted towards the complete sample of 6.7-GHz methanol masers detected in the MMB survey \citep{CasMMB10,GreenMMB10,CasMMB102,Green12}. Given the unbiased nature of the 6.7-GHz maser sample, our observations represent the most complete sample of 12.2-GHz methanol masers compiled to date. 

Over the presented longitude range, we detect 83 12.2-GHz methanol masers towards the 207 targeted 6.7-GHz methanol maser, a detection rate of 40 per cent. Our detections, combined with the two 12.2-GHz methanol masers detected by \citet{Caswell95b}, but not detected by us, bring the total number of 12.2-GHz masers in the 186$^{\circ}$~$<$~$l$~$<$~330$^{\circ}$, $b$ $=$ $\pm$2$^{\circ}$ to 85 (corresponding to a detection rate of 41 per cent). 39 of the observed 12.2-GHz methanol masers are new detections. The detection rate in the present longitude range is lower than the previously presented longitude range of 330$^{\circ}$ to 10$^{\circ}$ longitude range (40 compared to 46 per cent); attributable to the inclusion of the significantly higher detection rate in the 335$^{\circ}$ to 340$^{\circ}$ range \citep{Breen12stats}.

We give a list of 45 high luminosity 6.7-GHz methanol masers with 12.2-GHz counterparts that we believe are excellent candidates for follow-up observations of rarer methanol maser transitions, and in particular the transition at 107.0-GHz.

\section*{Acknowledgments}

The Parkes telescope is part of the Australia Telescope which is funded by the Commonwealth of Australia for operation as a National Facility managed by CSIRO. Financial support for this work was provided by the Australian
Research Council. This research has made use of: NASA's Astrophysics
Data System Abstract Service; and  the SIMBAD data base, operated at CDS, Strasbourg,
France.

\end{document}